\documentclass[%
 aip,
 amsmath,amssymb,
 reprint,%
]{revtex4-1}

\usepackage{graphicx}% Include figure files
\usepackage{dcolumn}% Align table columns on decimal point
\usepackage{bm}% bold math
\usepackage[utf8]{inputenc}
\usepackage[T1]{fontenc}
\usepackage{mathptmx}
\usepackage{etoolbox}
\usepackage[dvipsnames]{xcolor}
\usepackage{hyperref}

\makeatletter
\def\@email#1#2{%
 \endgroup
 \patchcmd{\titleblock@produce}
  {\frontmatter@RRAPformat}
  {\frontmatter@RRAPformat{\produce@RRAP{*#1\href{mailto:#2}{#2}}}\frontmatter@RRAPformat}
  {}{}
}%
\makeatother

\begin{document}

\preprint{AIP/123-QED}

\title{Thermal activation of low-density Ga implanted in Ge}

\author{Natalie D. Foster}
  \email{nfoster@utexas.edu}
 \affiliation{Sandia National Laboratories, Albuquerque, New Mexico 87185, USA}
 \affiliation{Department of Physics, The University of Texas at Austin, Austin, Texas 78712, USA}

\author{Andrew J. Miller}%
\affiliation{Sandia National Laboratories, Albuquerque, New Mexico 87185, USA}%
\author{Troy A. Hutchins-Delgado}%
\affiliation{Sandia National Laboratories, Albuquerque, New Mexico 87185, USA}%
\author{Christopher M. Smyth}
\affiliation{Sandia National Laboratories, Albuquerque, New Mexico 87185, USA}
\author{Michael C. Wanke}%
\affiliation{Sandia National Laboratories, Albuquerque, New Mexico 87185, USA}%
\author{Tzu-Ming Lu}
\affiliation{Sandia National Laboratories, Albuquerque, New Mexico 87185, USA}
\affiliation{Center for Integrated Nanotechnologies, Sandia National Laboratories, Albuquerque, New Mexico 87123, USA}
\author{Dwight R. Luhman}
\affiliation{Sandia National Laboratories, Albuquerque, New Mexico 87185, USA}

\date{\today}% It is always \today, today,
             %  but any date may be explicitly specified

\begin{abstract}
The nuclear spins of low-density implanted Ga atoms in Ge are interesting candidates for solid state-based qubits. To date, activation studies of implanted Ga in Ge have focused on high densities. Here we extend activation studies into the low-density regime. We use spreading resistance profiling and secondary ion mass spectrometry to derive electrical activation of Ga ions implanted into Ge as a function of rapid thermal anneal temperature and implant density. We show that for our implant conditions the activation is best for anneal temperatures between 400 and 650 $^\circ$C, with a maximum activation of 64\% at the highest fluence. Below 400 $^\circ$C, remaining implant damage results in defects that act as superfluous carriers, and above 650 $^\circ$C, surface roughening and loss of Ga ions are observed. The activation increased monotonically from 10\% to 64\% as the implant fluence increased from $6\times10^{10}$ to $6\times10^{12}$ cm$^{-2}$.  The results provide thermal anneal conditions to be used for initial studies of using low-density Ga atoms in Ge as nuclear spin qubits. 
\end{abstract}

\maketitle

% -------------------------- INTRODUCTION --------------------------------%
\section{\label{sec:intro}Introduction}
Nuclear spins in solid-state systems are compelling candidates for use as qubits. The concept of using donor nuclear spins as qubits in isotopically enriched silicon, originally proposed by Kane,\cite{Kane1998} is bolstered by the observations of long coherence times~\cite{Steger2013,Saeedi2013} even in integrated single donor devices.\cite{Muhonen2014} In silicon-based donor-quantum dot systems, an electron can be moved on and off the donor using voltages applied to lithographically defined surface electrodes in a lateral quantum dot geometry, creating a heterogenous qubit pair where the electron and nuclear spins interact via the hyperfine interaction.\cite{Muhonen2014} Entanglement between multiple nuclear spins through a simultaneously shared electron has even been demonstrated using this technique.\cite{Madzik2022} The promise of nuclear spin qubits has been the inspiration for a variety of approaches using donors\cite{Harvey-Collard2017,Morse2017,Deabreu2019} and isoelectronic spins\cite{Hensen2020} in silicon. 

While much research has focused on electron qubits utilizing quantum dots in silicon, holes in Ge quantum dots have also emerged as promising qubits. Recent advances in the growth of  high-quality planar Ge/SiGe heterostructures\cite{Laroche2016,Sammak2019} has led to the rapid development in just a few years of lateral quantum dots in Ge/SiGe,\cite{Hardy2019,Hendrickx2018} single hole qubit demonstrations,\cite{Hendrickx2020} and a four-hole-qubit device.\cite{Hendrickx2021} Hole spin qubits in Ge/SiGe, in contrast to electron spin qubits, have the capability of all-electric control of the hole spin through strong intrinsic spin orbit coupling,\cite{Hendrickx2020a} and exhibit tunable,\cite{Hendrickx2020} anisotropic g-factors.\cite{Miller2021} Together these features open the possibility for simplified and flexible control features in the qubit platform. 

The swift and promising emergence of hole-based qubits in Ge/SiGe quantum dots suggests that acceptor-based qubits in germanium also warrant attention. However, in contrast to donors in silicon, acceptor-based qubits have received less attention in the literature.\cite{Abadillo-Uriel2016,Kobayashi2021} Gallium is one candidate for an acceptor qubit in Ge. It has two stable isotopes at reasonable abundance, $^{69}$Ga (60.11\%) and $^{71}$Ga (39.89\%); both isotopes have nuclear spin-3/2. The $>$1/2 nuclear spin of the Ga isotopes could allow an individual spin to be controlled through nuclear electric resonance by leveraging the quadrupole moment as was recently demonstrated with $^{123}$Sb in silicon.\cite{Asaad2020} This method avoids the need for oscillating magnetic fields which places constraints on device design. All-electrical control of both the hole spin and nuclear spin in a qubit system would be highly desirable. 

An early step to exploring the possibility of using Ga as a nuclear spin qubit is to develop a reliable process for incorporating Ga atoms into substitutional sites of the Ge host crystal at the appropriate densities. Ga atoms can be introduced into a Ge host through growth, thermal diffusion, or ion implantation. Ion implantation is a standard technique for dopant incorporation that allows for great flexibility and is the focus of this work.  Following ion implantation, a thermal annealing step is typically required to repair implant damage and activate the implanted Ga atoms by incorporating them into substitutional sites. The activation ratio, defined in this work as the integrated 2D density of charge carriers to the integrated 2D density of implanted Ga atoms, is typically expected to increase with increasing annealing temperature owing to the higher thermal energy available. However, it is also known that at annealing temperatures too high, long range diffusion of dopants may occur, resulting in changes in the dopant distribution profiles and possibly a detrimental loss of dopants through the surface.\cite{Satta2006,Murrell2000} The goal is to find a tradeoff between dopant activation and dopant loss to achieve the appropriate annealing temperature window.

% ---------------------------- FIG 1 --------------------------------%
\begin{figure*}[t]
\includegraphics[width=0.8\textwidth]{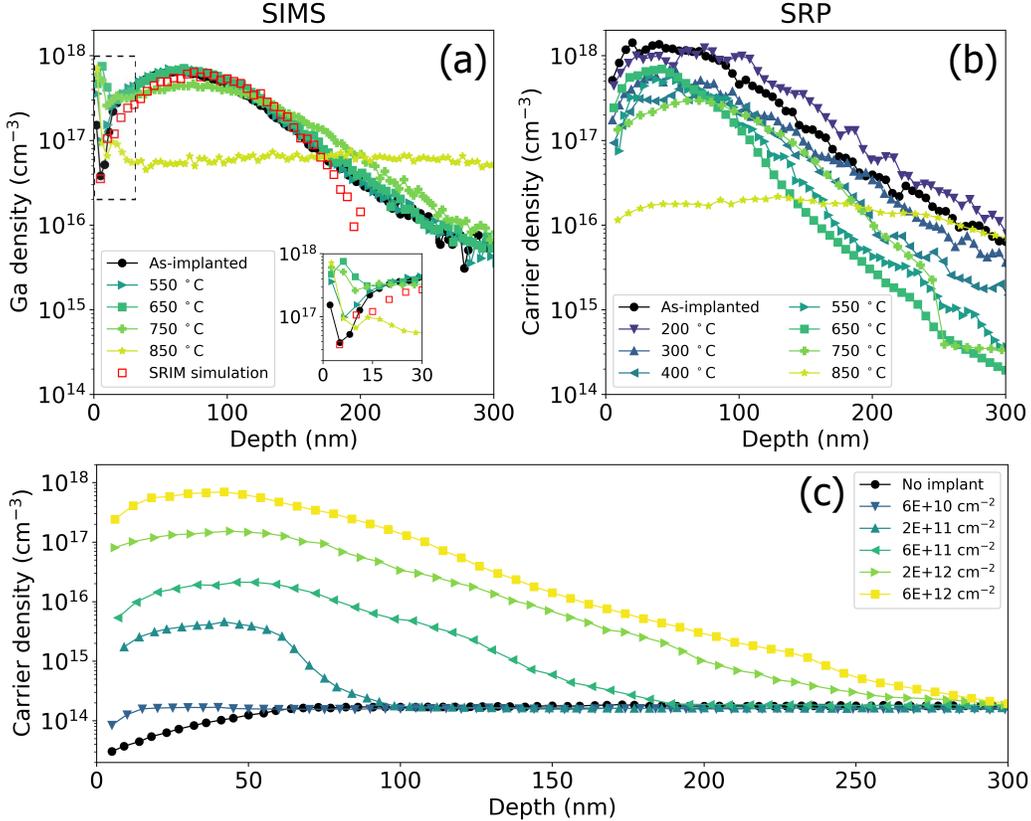}
\caption{\label{fig:depth_profiles}(a) SIMS-derived implanted Ga density as a function of depth for different anneal temperatures. SRIM simulation results are shown as red squares. The inset focuses on a smaller region near the surface outlined in a dashed box. (b) SRP-derived carrier density as a function of depth for different anneal temperatures. (c) SRP-derived carrier density for different implant fluences, including a sample that was not implanted. (a) and (b) are shown for fixed implant fluence of $6\times10^{12}$ cm$^{-2}$ and (c) for fixed anneal temperature 650 $^\circ$C.}
\end{figure*}

Activation of Ga-implanted Ge has been previously characterized for high density ion implantation ($>10^{20}$ cm$^{-2}$).\cite{Satta2006,Impellizzeri2009,Heera2010,Hellings2009} Lower fluences ($\leq 10^{12}$ cm$^{-2}$) are commonly used for single dopant spin qubit applications to facilitate individual dopant addressability and minimize interactions between neighbor dopants.\cite{Morello2010,Tracy2013} Activation of implanted dopants strongly depends on the implantation energy and fluence.  Thus, these prior studies on dopant activation in the high-fluence regime do not directly translate to the low-fluence regime required for acceptor qubit platforms.  Here, we present a systematic study of the activation of implanted Ga acceptors in a Ge substrate as a function of implant density in a low implant fluence regime and rapid thermal anneal (RTA) temperature with the goal of finding appropriate conditions for electrical activation of Ga in Ge.

% ---------------------------- FIG 2 --------------------------------%
\begin{figure*}[t]
\includegraphics[width=0.9\textwidth]{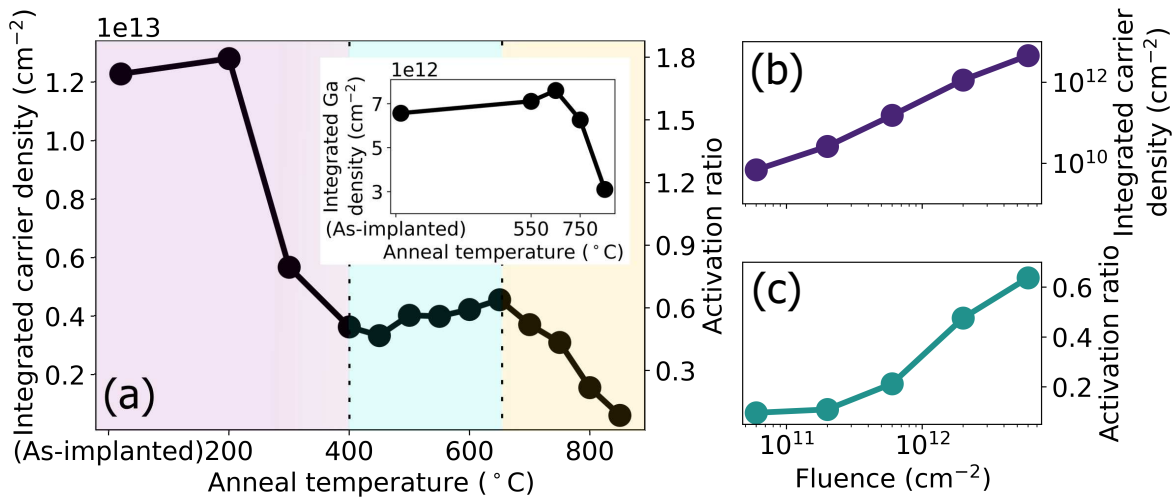}
\caption{\label{fig:activation} (a) Left y-axis: SRP-derived integrated carrier density. Right y-axis: corresponding activation ratio. The shaded regions divided by dashed vertical lines correspond to anneal temperature ranges exhibiting different effects from Ga dopant behavior: purple denotes unphysical activation ratio suggestive of defects, blue denotes stable electrical activation, and yellow denotes loss of dopants. Inset: SIMS-derived integrated Ga density. All data points shown for fixed implant fluence of $6\times10^{12}$ cm$^{-2}$. (b) SRP-derived integrated carrier density as a function of implant fluence. (c) Corresponding activation ratio for (b). All data points in (b) and (c) are for samples annealed at 650 $^\circ$C.}
\end{figure*}

% -------------------------- METHODS --------------------------------%
\section{\label{sec:methods}Methods}
We first simulated implanted dopant profiles as a function of implant energy and fluence prior to implantation using the Stopping and Range of Ions in Matter (SRIM) package.\cite{Ziegler2010} Based on the simulated implanted dopant profiles, we chose a beam energy of 175 keV and a maximum fluence of $6\times10^{12}$ cm$^{-2}$.  This combination spreads the Ga ions over a large enough 3D volume to keep the peak volume density below $\sim$10$^{18}$ cm$^{-3}$ where dopant clustering may occur which is not desirable for single dopant qubits.\cite{Sumikura2011}  Five (100)-oriented commercially acquired Ge wafers of low background doping ($\sim$10$^{12}$ cm$^{-3}$) were implanted with Ga ions by Kroko Inc. All implants were performed with a 7$^\circ$ incidence angle to minimize channeling. The Ga fluences for the 5 wafers were $6\times10^{10}$, $2\times10^{11}$, $6\times10^{11}$, $2\times10^{12}$, and $6\times10^{12}$ cm$^{-2}$, respectively, and were chosen to yield peak Ga densities between $6\times10^{15}$ cm$^{-3}$ and $6\times10^{17}$ cm$^{-3}$ . After implantation the wafers were diced using an automated dicing tool (MicroAuto). Then, the die were annealed in an RTA chamber (Jepelec) for 30 minutes in argon at atmospheric pressure. We explored different anneal temperatures ranging from 200 to 850 $^\circ$C. This range starts below the minimum temperature required to recrystallize damaged Ge~\cite{Hellings2009} and exceeds the temperature above which the onset of dopant loss is observed.\cite{Poon2005,Ioannou2008}

After the anneal step, selected die were analyzed by spreading resistance profiling (SRP, by Solecon) and secondary ion mass spectrometry (SIMS, by EAG), Raman spectroscopy, and atomic force microscopy (AFM). The SRP characterization determines electrically activated carrier density as a function of depth but does not differentiate between carriers due to activated implanted Ga dopants, preexisting impurities, or damage induced doping.  The SIMS characterization evaluates the Ga depth profile but does not differentiate between activated and non-activated Ga atoms.  We use these two complementary techniques to deduce the activation of Ga in Ge. Implant-induced defects in Ge, which are healed by annealing, were qualitatively evaluated as a function of implant and RTA conditions using Raman spectroscopy. Raman spectra were collected using a WITec 300a Raman spectrometer with an excitation wavelength of 532 nm, 18 mW laser power, 600 mm$^{-1}$ grating, 2.0 second integration time, 30 accumulations, and a 100$\times$ objective lens. Atomic force microscopy (Digital Instruments Veeco) was used to characterize the surface roughness as a function of the anneal temperature. For the AFM study, an implanted sample with fluence $6\times10^{12}$ cm$^{-2}$ was thermally annealed six times at incremental increasing temperatures between 600 and 850 $^\circ$C using the same recipes as the main series of samples. Using a surface defect as a registration mark, we acquired AFM and optical microscope images of the same area on the die prior to the first anneal and in between each of the anneal steps to track surface roughness evolution with annealing.

% ---------------------------- FIG 3 --------------------------------%
\begin{figure}[t]
\includegraphics[width=0.99\linewidth]{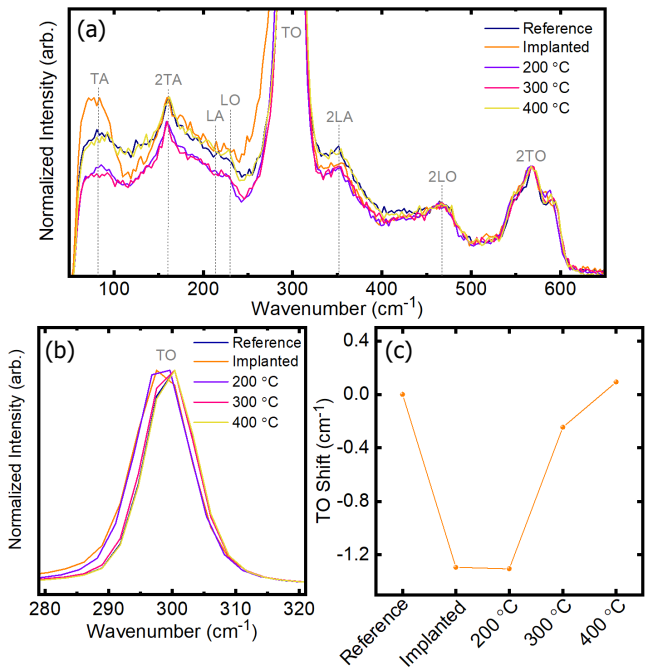}
\caption{\label{fig:raman} (a) Raman spectra normalized to the 2$^\mathrm{nd}$-order TO phonon mode zoomed in vertically to show evolution of low-intensity Ge Raman modes. (b) Normalized TO phonon mode spectra. (c) TO phonon mode shift. Data shown here were obtained from a Ge reference, after implanting with a Ga fluence of $6\times10^{12}$ cm$^{-2}$, and after 200 $^\circ$C, 300 $^\circ$C, and 400 $^\circ$C anneal.}
\end{figure}

\section{\label{sec:results}Results and Discussion}
Depth profile distributions of implanted Ga ions were used to evaluate the proportion of activated Ga dopants. Fig.~\ref{fig:depth_profiles}(a) shows the SIMS depth profiles of Ga in five different die, each implanted with a Ga fluence of $6\times10^{12}$ cm$^{-2}$. One die was not annealed (as-implanted) and the remaining four were annealed at temperatures of 550, 650, 750, and 850 $^\circ$C, respectively. The red squares represent density profiles from a SRIM simulation and agree reasonably well with the as-implanted SIMS density profile between 0 and 180 nm of the surface. As the RTA temperature increases, the SIMS depth profile remains roughly the same until 750 $^\circ$C at which point a slight broadening of the peak and reduction of the peak height are observable.  At 850 $^\circ$C the distribution becomes flat, indicating significant dopant diffusion and, as seen in the inset to Fig.~\ref{fig:activation}(a), concurrent dopant loss. The increase in the Ga density deeper than $\sim$180 nm for the sample annealed at 850 $^\circ$C indicates that some dopants are diffusing deeper into the substrate while the inset of Fig.~\ref{fig:depth_profiles}(a), which displays the close-up of profiles near the surface, indicates that there is also significant diffusion toward the surface and segregation.  Ga dopants appear to accumulate more at the surface as the RTA temperature increases.

Fig.~\ref{fig:depth_profiles}(b) shows electrical carrier density profiles derived from SRP. This figure includes data from the same samples as in Fig.~\ref{fig:depth_profiles}(a), plus data from a series of samples annealed at lower temperatures. SRP profiles gradually narrow as the anneal temperature increases up to 650 $^\circ$C, and then rapidly broaden and decrease in amplitude for increasing temperatures above 650 $^\circ$C.  The broadening at high temperatures is consistent with the 750 and 850 $^\circ$C data shown in Fig.~\ref{fig:depth_profiles}(a) for SIMS analysis, suggesting rapid dopant diffusion and loss of dopants at high RTA temperatures.\cite{Chroneos2014}
To further understand the carrier and Ga distribution dependence on the anneal temperature, we plot derived integrated carrier density (left axis) and corresponding activation ratio (right axis) as a function of anneal temperature for samples implanted with fluence $6\times10^{12}$ cm$^{-2}$ in Fig.~\ref{fig:activation}(a). Here we define activation ratio for each die as the total number of charge carriers derived from SRP measurements divided by the total number of Ga ions derived from SIMS measurements on the particular die annealed at 550 $^\circ$C. The figure is divided into three shaded regions by anneal temperature: i) $T < 400$ $^\circ$C (purple), ii) $400 < T < 650$ $^\circ$C (blue), and iii) $T > 650$ $^\circ$C (yellow). 
In the first region we observe a large and unphysical (i.e $>1$) activation ratio. Since we defined the activation ratio as the number of free carriers divided by the number of implanted ions, additional carriers must be invoked to explain the unphysical activation ratios $>1$ in the purple shaded region of Fig.~\ref{fig:activation}(a). 

The unphysical activation ratio and narrowing distributions in Fig.~\ref{fig:depth_profiles}(b) at lower anneal temperatures are likely an artifact due to damage induced doping. We used Raman spectroscopy as a probe of implant-induced defects.  Raman spectroscopy was carried out on unimplanted, as-implanted, and annealed samples. The implanted samples received a fluence of $6\times10^{12}$ cm$^{-2}$. The annealed samples were annealed at 200, 300, and 400 $^\circ$C. Fig.~\ref{fig:raman}(a) shows annotated Raman spectra normalized to the 2$^\mathrm{nd}$-order transverse optical (TO) phonon mode of Ge to clearly show trends in the intensities and line shapes of various peaks in the Ge phonon spectrum as a function of process conditions. We first note that the intensity of the transverse acoustic (TA) mode ($\sim$85 cm$^{-1}$) increases relative to that of optical phonon modes after implant.  This is caused by the symmetry breaking of the primary acoustic mode with implant-induced disorder in the Ge lattice.\cite{Lazarenkova2003} The peak wavenumber and line shape of the Ge TO mode ($\sim$300 cm$^{-1}$) provides a proxy for the degree of disorder and strain in the lattice.\cite{Pizani2000}
Fig.~\ref{fig:raman}(b) shows a narrow spectrum highlighting the prominent TO phonon mode of Ge near 300 cm$^{-1}$. Fig.~\ref{fig:raman}(c) summarizes the TO mode shift derived from Fig.~\ref{fig:raman}(b) with respect to a reference bulk Ge sample. The TO mode exhibits a 1.3 cm$^{-1}$ decrease in wavenumber after implantation due to the dampened restoring force for the corresponding lattice vibration and tensile strain imparted on the lattice~\cite{Cheng2013,Sui1993} by the increased defect concentration after the implantation. The sudden shift of the TO mode back towards the reference value with increasing anneal temperature above 200 $^\circ$C indicates the RTA heals crystalline disorder and relieves tensile strain.\cite{Yukhymchyk2015} The TO mode shift at 400 $^\circ$C appears to be the same as for an unimplanted sample, suggesting that the Ge lattice is healed by annealing at 400 $^\circ$C. We assume that the lattice healed when annealed at even higher temperatures as well. Therefore, we attribute the large carrier density and unphysical activation ratio $>1$ for anneal temperatures below 400 $^\circ$C in the purple region of Fig.~\ref{fig:activation}(a) to free carriers in the damaged lattice while we consider the carrier density measured for anneal temperatures above 400 $^\circ$C to accurately represent the number of activated Ga ions.

In the yellow region of Fig.~\ref{fig:activation}(a) ($T>650$ $^\circ$C) we observe a monotonic decrease in activation as the anneal temperature increases. The inset of Fig.~\ref{fig:activation}(a) shows the integrated Ga areal density from SIMS as a function of anneal temperature.  As the SIMS technique does not differentiate activated and non-activated Ga atoms, the observed activation decrease for $T > 650$ $^\circ$C indicates a loss of Ga ions in samples annealed at the highest two temperatures. This observation is consistent with a previous report showing 50\% dopant loss in crystalline germanium after annealing at 800 $^\circ$C compared with 700 $^\circ$C.\cite{Lieten2013}  Given that our definition for the activation ratio normalizes the number of measured carriers by the number of implanted ions rather than by the number of ions remaining after annealing, we can attribute the steadily decreasing activation in the yellow region of Fig.~\ref{fig:activation}(a) to the loss of dopants.

\begin{figure}[t]
\includegraphics[width=0.95\linewidth]{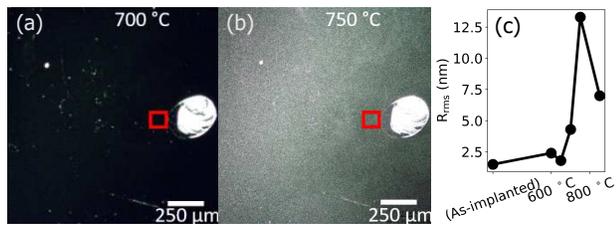}
\caption{\label{fig:AFM} Dark field images of sample surface following (a) 700 and (b) 750 $^\circ$C annealing. The red box denotes the AFM scanning area. The large white oval is a surface defect used for image registration (c) RMS roughness acquired with AFM over the area in the red box as a function of anneal temperature. The sample has been implanted with fluence of $6\times10^{12}$ cm$^{-2}$.}
\end{figure}

To further explore the loss of dopants to the surface at high anneal temperature, as indicated by the SIMS data in the inset of Fig.~\ref{fig:activation}(a), we use insight from AFM surface analysis. As described in the fabrication section, the sample was annealed multiple times, and dark field optical images of the sample after the 700 $^\circ$C and  750 $^\circ$C annealing steps are shown in Figs.~\ref{fig:AFM}(a) and \ref{fig:AFM}(b), respectively. The figures show qualitative surface changes between anneal steps and the 225 $\mu$m$^2$ AFM scan area (outlined in red) is positioned near a large surface feature for ease of consistency between scans (white spot, far right). Fig.~\ref{fig:AFM}(c) shows root-mean-square (rms) roughness vs. anneal temperature with the appearance of an abrupt increase in roughness at 750 $^\circ$C followed by a decrease for $T > 750$ $^\circ$C. Temperatures corresponding to increased roughness ($T>700$ $^\circ$C) fall within the yellow shaded region of Fig.~\ref{fig:activation}(a), where activation steadily decreases as the anneal temperature increases. Previous reports have showed surface roughening is accompanied by dopant loss,\cite{Suh2005,Poon2005,Ioannou2008} which supports our hypothesis that significant diffusion and loss of dopants through the surface occurs for high anneal temperatures $>$700 $^\circ$C . Our observations of dopant loss and surface roughening in Ga-implanted Ge are consistent with these prior reports.  

The combination of the observed activation ratio, Raman characterization and AFM surface studies suggests a usable range of thermal anneal temperatures of $400 < T < 650$ $^\circ$C for electrical activation of Ga dopants in Ge. This range is represented by the blue (center) shaded region of Fig.~\ref{fig:activation}(a).  Above 400 $^\circ$C, the structural defects and strain induced by the implant process itself appears to be removed so that the measured carrier density can be attributed to the Ga ions. Below 650 $^\circ$C, the surface remains smooth, and surface segregation, Ga diffusion and Ga loss appear negligible. Previous activation studies of Ga-doped Ge at higher densities~\cite{Hellings2009,Ioannou2008} observed negligible diffusion of dopants for anneal temperatures between 400 $^\circ$C and 700 $^\circ$C, consistent with our findings. In this region, activation is reasonable and increases slowly with temperature, which is expected due to the higher temperature enabling more Ga to move onto substitutional sites. The relatively small change in activation ratio versus anneal temperature in this region suggests that the majority of Ga atoms that can move onto substitutional sites have done so.  

% ---------------------------- FIG 4 --------------------------------%

Having identified an appropriate temperature range for the implant anneal, we turn our attention to the dependence of the activation ratio on implant fluence. Since the activation ratio in this range increases with temperature, we chose the highest temperature within the range (650 $^\circ$C) to explore the fluence dependence.  Fig.~\ref{fig:depth_profiles}(c) shows the carrier density profiles derived from SRP for each of the 5 implanted wafers with implant fluences ranging from $6\times10^{10}$ to $6\times10^{12}$ cm$^{-2}$ for a fixed anneal temperature of 650 $^\circ$C, and for an unimplanted Ge wafer used as a control. The peak carrier density depth around 50 nm, appears to be independent of the fluence.  However, the amplitude of the peak carrier density depends strongly on the fluence and increases faster than the implant density. Fig.~\ref{fig:activation}(b) shows that the SRP-derived integrated carrier density increases superlinearly with the fluence indicating activation improves as the fluence increases. This leads to Fig.~\ref{fig:activation}(c), which shows that the activation ratio increases monotonically from 10\% to 64\% as the fluence increases from $6\times10^{10}$ to $6\times10^{12}$ cm$^{-2}$.  The monotonic increase in the activation ratio with fluence suggests that a larger fraction of Ga dopants are able to incorporate into substitutional lattice sites with increasing fluence without suffering amorphization effects known to plague implanted germanium at high implant fluences.\cite{Koffel2009}  The highest activation of 64\% is comparable to typical measured values from donors implanted in silicon.\cite{Holmes2019} In the context of nuclear spin-based qubits, dopants implanted with fluences in the range of those studied in this work yielded successful implementations of individual donor spin qubits in Si,\cite{Asaad2020,Pla2012,Pla2013} an encouraging figure for future work targeting acceptor spin qubits in Ge.

\section{\label{sec:conclusion}Conclusion}
The dopant activation ratios of a series of high purity Ga-implanted Ge wafers were examined as a function of thermal anneal temperature and implant fluence using depth profile measurements and surface analysis studies. We found that the usable annealing temperature range for electrical activation of implanted dopants lies between 400 and 650 $^\circ$C. At lower temperatures, damage from the implant process appears to result in unwanted excess free carriers, while at higher temperatures dopants are lost via diffusion and surface quality degradation is observed. Additionally, the activation ratio increases with increased implant fluence, and a high degree of electrical activation (near 64\%) was obtained at our highest fluence of $6\times10^{12}$ cm$^{-2}$. The insights presented here offer a guide towards improving electrical activation of Ga dopants in Ge for eventual studies of all-electric control of acceptor nuclear spin qubits in Ge.

\begin{acknowledgments}
This research was primarily supported by the Laboratory Directed Research and Development Program at Sandia National Laboratories (SNL). This work was performed, in part, at the Center for Integrated Nanotechnologies, an Office of Science User Facility operated for the U.S. Department of Energy (DOE) Office of Science. SNL is a multi-mission laboratory managed and operated by National Technology and Engineering Solutions of Sandia LLC, a wholly owned subsidiary of Honeywell International Inc. for the U.S. DOE National Nuclear Security Administration under contract DE-NA0003525. This paper describes objective technical results and analysis. Any subjective views or opinions that might be expressed in the paper do not necessarily represent the views of the U.S. DOE or the United States Government. The data that support the findings of this study are available from the corresponding author upon reasonable request. N. D. F. acknowledges additional support from NSF GRFP DGE-1610403.
\end{acknowledgments}

%\nocite{*}
\bibliography{Ga-Ge_activation}

%merlin.mbs aipnum4-1.bst 2010-07-25 4.21a (PWD, AO, DPC) hacked
%Control: key (0)
%Control: author (8) initials jnrlst
%Control: editor formatted (1) identically to author
%Control: production of article title (0) allowed
%Control: page (1) range
%Control: year (1) truncated
%Control: production of eprint (0) enabled
\begin{thebibliography}{43}%
\makeatletter
\providecommand \@ifxundefined [1]{%
 \@ifx{#1\undefined}
}%
\providecommand \@ifnum [1]{%
 \ifnum #1\expandafter \@firstoftwo
 \else \expandafter \@secondoftwo
 \fi
}%
\providecommand \@ifx [1]{%
 \ifx #1\expandafter \@firstoftwo
 \else \expandafter \@secondoftwo
 \fi
}%
\providecommand \natexlab [1]{#1}%
\providecommand \enquote  [1]{``#1''}%
\providecommand \bibnamefont  [1]{#1}%
\providecommand \bibfnamefont [1]{#1}%
\providecommand \citenamefont [1]{#1}%
\providecommand \href@noop [0]{\@secondoftwo}%
\providecommand \href [0]{\begingroup \@sanitize@url \@href}%
\providecommand \@href[1]{\@@startlink{#1}\@@href}%
\providecommand \@@href[1]{\endgroup#1\@@endlink}%
\providecommand \@sanitize@url [0]{\catcode `\\12\catcode `\$12\catcode
  `\&12\catcode `\#12\catcode `\^12\catcode `\_12\catcode `\%12\relax}%
\providecommand \@@startlink[1]{}%
\providecommand \@@endlink[0]{}%
\providecommand \url  [0]{\begingroup\@sanitize@url \@url }%
\providecommand \@url [1]{\endgroup\@href {#1}{\urlprefix }}%
\providecommand \urlprefix  [0]{URL }%
\providecommand \Eprint [0]{\href }%
\providecommand \doibase [0]{http://dx.doi.org/}%
\providecommand \selectlanguage [0]{\@gobble}%
\providecommand \bibinfo  [0]{\@secondoftwo}%
\providecommand \bibfield  [0]{\@secondoftwo}%
\providecommand \translation [1]{[#1]}%
\providecommand \BibitemOpen [0]{}%
\providecommand \bibitemStop [0]{}%
\providecommand \bibitemNoStop [0]{.\EOS\space}%
\providecommand \EOS [0]{\spacefactor3000\relax}%
\providecommand \BibitemShut  [1]{\csname bibitem#1\endcsname}%
\let\auto@bib@innerbib\@empty
%</preamble>
\bibitem [{\citenamefont {Kane}(1998)}]{Kane1998}%
  \BibitemOpen
  \bibfield  {author} {\bibinfo {author} {\bibfnamefont {B.~E.}\ \bibnamefont
  {Kane}},\ }\bibfield  {title} {\enquote {\bibinfo {title} {{A silicon-based
  nuclear spin quantum computer}},}\ }\href {\doibase
  10.1142/9789812701633_0003} {\bibfield  {journal} {\bibinfo  {journal}
  {Nature}\ }\textbf {\bibinfo {volume} {393}},\ \bibinfo {pages} {133--137}
  (\bibinfo {year} {1998})}\BibitemShut {NoStop}%
\bibitem [{\citenamefont {Steger}\ \emph {et~al.}(2012)\citenamefont {Steger},
  \citenamefont {Saeedi}, \citenamefont {Thewalt}, \citenamefont {Morton},
  \citenamefont {Riemann}, \citenamefont {Abrosimov}, \citenamefont {Becker},\
  and\ \citenamefont {Pohl}}]{Steger2013}%
  \BibitemOpen
  \bibfield  {author} {\bibinfo {author} {\bibfnamefont {M.}~\bibnamefont
  {Steger}}, \bibinfo {author} {\bibfnamefont {K.}~\bibnamefont {Saeedi}},
  \bibinfo {author} {\bibfnamefont {M.~L.~W.}\ \bibnamefont {Thewalt}},
  \bibinfo {author} {\bibfnamefont {J.~J.~L.}\ \bibnamefont {Morton}}, \bibinfo
  {author} {\bibfnamefont {H.}~\bibnamefont {Riemann}}, \bibinfo {author}
  {\bibfnamefont {N.~V.}\ \bibnamefont {Abrosimov}}, \bibinfo {author}
  {\bibfnamefont {P.}~\bibnamefont {Becker}}, \ and\ \bibinfo {author}
  {\bibfnamefont {H.-J.}\ \bibnamefont {Pohl}},\ }\bibfield  {title} {\enquote
  {\bibinfo {title} {{Quantum Information Storage for over 180 s Using Donor
  Spins in a 28Si "Semiconductor Vacuum"}},}\ }\href {\doibase
  10.1126/science.1217635} {\bibfield  {journal} {\bibinfo  {journal}
  {Science}\ }\textbf {\bibinfo {volume} {336}},\ \bibinfo {pages} {1280--1284}
  (\bibinfo {year} {2012})}\BibitemShut {NoStop}%
\bibitem [{\citenamefont {Saeedi}\ \emph {et~al.}(2013)\citenamefont {Saeedi},
  \citenamefont {Simmons}, \citenamefont {Salvail}, \citenamefont {Dluhy},
  \citenamefont {Riemann}, \citenamefont {Abrosimov}, \citenamefont {Becker},
  \citenamefont {Pohl}, \citenamefont {Morton},\ and\ \citenamefont
  {Thewalt}}]{Saeedi2013}%
  \BibitemOpen
  \bibfield  {author} {\bibinfo {author} {\bibfnamefont {K.}~\bibnamefont
  {Saeedi}}, \bibinfo {author} {\bibfnamefont {S.}~\bibnamefont {Simmons}},
  \bibinfo {author} {\bibfnamefont {J.~Z.}\ \bibnamefont {Salvail}}, \bibinfo
  {author} {\bibfnamefont {P.}~\bibnamefont {Dluhy}}, \bibinfo {author}
  {\bibfnamefont {H.}~\bibnamefont {Riemann}}, \bibinfo {author} {\bibfnamefont
  {N.~V.}\ \bibnamefont {Abrosimov}}, \bibinfo {author} {\bibfnamefont
  {P.}~\bibnamefont {Becker}}, \bibinfo {author} {\bibfnamefont {H.~J.}\
  \bibnamefont {Pohl}}, \bibinfo {author} {\bibfnamefont {J.~J.}\ \bibnamefont
  {Morton}}, \ and\ \bibinfo {author} {\bibfnamefont {M.~L.}\ \bibnamefont
  {Thewalt}},\ }\bibfield  {title} {\enquote {\bibinfo {title}
  {{Room-temperature quantum bit storage exceeding 39 minutes using ionized
  donors in silicon-28}},}\ }\href {\doibase 10.1126/science.1239584}
  {\bibfield  {journal} {\bibinfo  {journal} {Science}\ }\textbf {\bibinfo
  {volume} {342}},\ \bibinfo {pages} {830--833} (\bibinfo {year}
  {2013})}\BibitemShut {NoStop}%
\bibitem [{\citenamefont {Muhonen}\ \emph {et~al.}(2014)\citenamefont
  {Muhonen}, \citenamefont {Dehollain}, \citenamefont {Laucht}, \citenamefont
  {Hudson}, \citenamefont {Kalra}, \citenamefont {Sekiguchi}, \citenamefont
  {Itoh}, \citenamefont {Jamieson}, \citenamefont {McCallum}, \citenamefont
  {Dzurak},\ and\ \citenamefont {Morello}}]{Muhonen2014}%
  \BibitemOpen
  \bibfield  {author} {\bibinfo {author} {\bibfnamefont {J.~T.}\ \bibnamefont
  {Muhonen}}, \bibinfo {author} {\bibfnamefont {J.~P.}\ \bibnamefont
  {Dehollain}}, \bibinfo {author} {\bibfnamefont {A.}~\bibnamefont {Laucht}},
  \bibinfo {author} {\bibfnamefont {F.~E.}\ \bibnamefont {Hudson}}, \bibinfo
  {author} {\bibfnamefont {R.}~\bibnamefont {Kalra}}, \bibinfo {author}
  {\bibfnamefont {T.}~\bibnamefont {Sekiguchi}}, \bibinfo {author}
  {\bibfnamefont {K.~M.}\ \bibnamefont {Itoh}}, \bibinfo {author}
  {\bibfnamefont {D.~N.}\ \bibnamefont {Jamieson}}, \bibinfo {author}
  {\bibfnamefont {J.~C.}\ \bibnamefont {McCallum}}, \bibinfo {author}
  {\bibfnamefont {A.~S.}\ \bibnamefont {Dzurak}}, \ and\ \bibinfo {author}
  {\bibfnamefont {A.}~\bibnamefont {Morello}},\ }\bibfield  {title} {\enquote
  {\bibinfo {title} {{Storing quantum information for 30 seconds in a
  nanoelectronic device}},}\ }\href {\doibase 10.1038/nnano.2014.211}
  {\bibfield  {journal} {\bibinfo  {journal} {Nature Nanotechnology}\ }\textbf
  {\bibinfo {volume} {9}},\ \bibinfo {pages} {986--991} (\bibinfo {year}
  {2014})},\ \Eprint {http://arxiv.org/abs/1402.7140} {arXiv:1402.7140}
  \BibitemShut {NoStop}%
\bibitem [{\citenamefont {Madzik}\ \emph {et~al.}(2022)\citenamefont {Madzik},
  \citenamefont {Asaad}, \citenamefont {Youssry}, \citenamefont {Joecker},
  \citenamefont {Rudinger}, \citenamefont {Nielsen}, \citenamefont {Young},
  \citenamefont {Proctor}, \citenamefont {Baczewski}, \citenamefont {Laucht},
  \citenamefont {Schmitt}, \citenamefont {Hudson}, \citenamefont {Itoh},
  \citenamefont {Jakob}, \citenamefont {Johnson}, \citenamefont {Jamieson},
  \citenamefont {Dzurak}, \citenamefont {Ferrie}, \citenamefont
  {Blume-Kohout},\ and\ \citenamefont {Morello}}]{Madzik2022}%
  \BibitemOpen
  \bibfield  {author} {\bibinfo {author} {\bibfnamefont {M.~T.}\ \bibnamefont
  {Madzik}}, \bibinfo {author} {\bibfnamefont {S.}~\bibnamefont {Asaad}},
  \bibinfo {author} {\bibfnamefont {A.}~\bibnamefont {Youssry}}, \bibinfo
  {author} {\bibfnamefont {B.}~\bibnamefont {Joecker}}, \bibinfo {author}
  {\bibfnamefont {K.~M.}\ \bibnamefont {Rudinger}}, \bibinfo {author}
  {\bibfnamefont {E.}~\bibnamefont {Nielsen}}, \bibinfo {author} {\bibfnamefont
  {K.~C.}\ \bibnamefont {Young}}, \bibinfo {author} {\bibfnamefont {T.~J.}\
  \bibnamefont {Proctor}}, \bibinfo {author} {\bibfnamefont {A.~D.}\
  \bibnamefont {Baczewski}}, \bibinfo {author} {\bibfnamefont {A.}~\bibnamefont
  {Laucht}}, \bibinfo {author} {\bibfnamefont {V.}~\bibnamefont {Schmitt}},
  \bibinfo {author} {\bibfnamefont {F.~E.}\ \bibnamefont {Hudson}}, \bibinfo
  {author} {\bibfnamefont {K.~M.}\ \bibnamefont {Itoh}}, \bibinfo {author}
  {\bibfnamefont {A.~M.}\ \bibnamefont {Jakob}}, \bibinfo {author}
  {\bibfnamefont {B.~C.}\ \bibnamefont {Johnson}}, \bibinfo {author}
  {\bibfnamefont {D.~N.}\ \bibnamefont {Jamieson}}, \bibinfo {author}
  {\bibfnamefont {A.~S.}\ \bibnamefont {Dzurak}}, \bibinfo {author}
  {\bibfnamefont {C.}~\bibnamefont {Ferrie}}, \bibinfo {author} {\bibfnamefont
  {R.}~\bibnamefont {Blume-Kohout}}, \ and\ \bibinfo {author} {\bibfnamefont
  {A.}~\bibnamefont {Morello}},\ }\bibfield  {title} {\enquote {\bibinfo
  {title} {{Precision tomography of a three-qubit donor quantum processor in
  silicon}},}\ }\href {\doibase 10.1038/s41586-021-04292-7} {\bibfield
  {journal} {\bibinfo  {journal} {Nature}\ }\textbf {\bibinfo {volume} {601}},\
  \bibinfo {pages} {348--353} (\bibinfo {year} {2022})}\BibitemShut {NoStop}%
\bibitem [{\citenamefont {Harvey-Collard}\ \emph {et~al.}(2017)\citenamefont
  {Harvey-Collard}, \citenamefont {Jacobson}, \citenamefont {Rudolph},
  \citenamefont {Dominguez}, \citenamefont {{Ten Eyck}}, \citenamefont {Wendt},
  \citenamefont {Pluym}, \citenamefont {Gamble}, \citenamefont {Lilly},
  \citenamefont {Pioro-Ladri{\`{e}}re},\ and\ \citenamefont
  {Carroll}}]{Harvey-Collard2017}%
  \BibitemOpen
  \bibfield  {author} {\bibinfo {author} {\bibfnamefont {P.}~\bibnamefont
  {Harvey-Collard}}, \bibinfo {author} {\bibfnamefont {N.~T.}\ \bibnamefont
  {Jacobson}}, \bibinfo {author} {\bibfnamefont {M.}~\bibnamefont {Rudolph}},
  \bibinfo {author} {\bibfnamefont {J.}~\bibnamefont {Dominguez}}, \bibinfo
  {author} {\bibfnamefont {G.~A.}\ \bibnamefont {{Ten Eyck}}}, \bibinfo
  {author} {\bibfnamefont {J.~R.}\ \bibnamefont {Wendt}}, \bibinfo {author}
  {\bibfnamefont {T.}~\bibnamefont {Pluym}}, \bibinfo {author} {\bibfnamefont
  {J.~K.}\ \bibnamefont {Gamble}}, \bibinfo {author} {\bibfnamefont {M.~P.}\
  \bibnamefont {Lilly}}, \bibinfo {author} {\bibfnamefont {M.}~\bibnamefont
  {Pioro-Ladri{\`{e}}re}}, \ and\ \bibinfo {author} {\bibfnamefont {M.~S.}\
  \bibnamefont {Carroll}},\ }\bibfield  {title} {\enquote {\bibinfo {title}
  {{Coherent coupling between a quantum dot and a donor in silicon}},}\ }\href
  {\doibase 10.1038/s41467-017-01113-2} {\bibfield  {journal} {\bibinfo
  {journal} {Nature Communications}\ }\textbf {\bibinfo {volume} {8}},\
  \bibinfo {pages} {1--6} (\bibinfo {year} {2017})},\ \Eprint
  {http://arxiv.org/abs/1512.01606} {arXiv:1512.01606} \BibitemShut {NoStop}%
\bibitem [{\citenamefont {Morse}\ \emph {et~al.}(2017)\citenamefont {Morse},
  \citenamefont {Abraham}, \citenamefont {DeAbreu}, \citenamefont {Bowness},
  \citenamefont {Richards}, \citenamefont {Riemann}, \citenamefont {Abrosimov},
  \citenamefont {Becker}, \citenamefont {Pohl}, \citenamefont {Thewalt},\ and\
  \citenamefont {Simmons}}]{Morse2017}%
  \BibitemOpen
  \bibfield  {author} {\bibinfo {author} {\bibfnamefont {K.~J.}\ \bibnamefont
  {Morse}}, \bibinfo {author} {\bibfnamefont {R.~J.}\ \bibnamefont {Abraham}},
  \bibinfo {author} {\bibfnamefont {A.}~\bibnamefont {DeAbreu}}, \bibinfo
  {author} {\bibfnamefont {C.}~\bibnamefont {Bowness}}, \bibinfo {author}
  {\bibfnamefont {T.~S.}\ \bibnamefont {Richards}}, \bibinfo {author}
  {\bibfnamefont {H.}~\bibnamefont {Riemann}}, \bibinfo {author} {\bibfnamefont
  {N.~V.}\ \bibnamefont {Abrosimov}}, \bibinfo {author} {\bibfnamefont
  {P.}~\bibnamefont {Becker}}, \bibinfo {author} {\bibfnamefont {H.~J.}\
  \bibnamefont {Pohl}}, \bibinfo {author} {\bibfnamefont {M.~L.}\ \bibnamefont
  {Thewalt}}, \ and\ \bibinfo {author} {\bibfnamefont {S.}~\bibnamefont
  {Simmons}},\ }\bibfield  {title} {\enquote {\bibinfo {title} {{A photonic
  platform for donor spin qubits in silicon}},}\ }\href {\doibase
  10.1126/sciadv.1700930} {\bibfield  {journal} {\bibinfo  {journal} {Science
  Advances}\ }\textbf {\bibinfo {volume} {3}},\ \bibinfo {pages} {1--11}
  (\bibinfo {year} {2017})},\ \Eprint {http://arxiv.org/abs/1606.03488}
  {arXiv:1606.03488} \BibitemShut {NoStop}%
\bibitem [{\citenamefont {Deabreu}\ \emph {et~al.}(2019)\citenamefont
  {Deabreu}, \citenamefont {Bowness}, \citenamefont {Abraham}, \citenamefont
  {Medvedova}, \citenamefont {Morse}, \citenamefont {Riemann}, \citenamefont
  {Abrosimov}, \citenamefont {Becker}, \citenamefont {Pohl}, \citenamefont
  {Thewalt},\ and\ \citenamefont {Simmons}}]{Deabreu2019}%
  \BibitemOpen
  \bibfield  {author} {\bibinfo {author} {\bibfnamefont {A.}~\bibnamefont
  {Deabreu}}, \bibinfo {author} {\bibfnamefont {C.}~\bibnamefont {Bowness}},
  \bibinfo {author} {\bibfnamefont {R.~J.}\ \bibnamefont {Abraham}}, \bibinfo
  {author} {\bibfnamefont {A.}~\bibnamefont {Medvedova}}, \bibinfo {author}
  {\bibfnamefont {K.~J.}\ \bibnamefont {Morse}}, \bibinfo {author}
  {\bibfnamefont {H.}~\bibnamefont {Riemann}}, \bibinfo {author} {\bibfnamefont
  {N.~V.}\ \bibnamefont {Abrosimov}}, \bibinfo {author} {\bibfnamefont
  {P.}~\bibnamefont {Becker}}, \bibinfo {author} {\bibfnamefont {H.~J.}\
  \bibnamefont {Pohl}}, \bibinfo {author} {\bibfnamefont {M.~L.}\ \bibnamefont
  {Thewalt}}, \ and\ \bibinfo {author} {\bibfnamefont {S.}~\bibnamefont
  {Simmons}},\ }\bibfield  {title} {\enquote {\bibinfo {title}
  {{Characterization of the Si: Se+ Spin-Photon Interface}},}\ }\href {\doibase
  10.1103/PhysRevApplied.11.044036} {\bibfield  {journal} {\bibinfo  {journal}
  {Physical Review Applied}\ }\textbf {\bibinfo {volume} {11}},\ \bibinfo
  {pages} {1} (\bibinfo {year} {2019})},\ \Eprint
  {http://arxiv.org/abs/1809.10228} {arXiv:1809.10228} \BibitemShut {NoStop}%
\bibitem [{\citenamefont {Hensen}\ \emph {et~al.}(2020)\citenamefont {Hensen},
  \citenamefont {{Wei Huang}}, \citenamefont {Yang}, \citenamefont {{Wai
  Chan}}, \citenamefont {Yoneda}, \citenamefont {Tanttu}, \citenamefont
  {Hudson}, \citenamefont {Laucht}, \citenamefont {Itoh}, \citenamefont {Ladd},
  \citenamefont {Morello},\ and\ \citenamefont {Dzurak}}]{Hensen2020}%
  \BibitemOpen
  \bibfield  {author} {\bibinfo {author} {\bibfnamefont {B.}~\bibnamefont
  {Hensen}}, \bibinfo {author} {\bibfnamefont {W.}~\bibnamefont {{Wei Huang}}},
  \bibinfo {author} {\bibfnamefont {C.~H.}\ \bibnamefont {Yang}}, \bibinfo
  {author} {\bibfnamefont {K.}~\bibnamefont {{Wai Chan}}}, \bibinfo {author}
  {\bibfnamefont {J.}~\bibnamefont {Yoneda}}, \bibinfo {author} {\bibfnamefont
  {T.}~\bibnamefont {Tanttu}}, \bibinfo {author} {\bibfnamefont {F.~E.}\
  \bibnamefont {Hudson}}, \bibinfo {author} {\bibfnamefont {A.}~\bibnamefont
  {Laucht}}, \bibinfo {author} {\bibfnamefont {K.~M.}\ \bibnamefont {Itoh}},
  \bibinfo {author} {\bibfnamefont {T.~D.}\ \bibnamefont {Ladd}}, \bibinfo
  {author} {\bibfnamefont {A.}~\bibnamefont {Morello}}, \ and\ \bibinfo
  {author} {\bibfnamefont {A.~S.}\ \bibnamefont {Dzurak}},\ }\bibfield  {title}
  {\enquote {\bibinfo {title} {{A silicon quantum-dot-coupled nuclear spin
  qubit}},}\ }\href {\doibase 10.1038/s41565-019-0587-7} {\bibfield  {journal}
  {\bibinfo  {journal} {Nature Nanotechnology}\ }\textbf {\bibinfo {volume}
  {15}},\ \bibinfo {pages} {13--17} (\bibinfo {year} {2020})},\ \Eprint
  {http://arxiv.org/abs/1904.08260} {arXiv:1904.08260} \BibitemShut {NoStop}%
\bibitem [{\citenamefont {Laroche}\ \emph {et~al.}(2016)\citenamefont
  {Laroche}, \citenamefont {Huang}, \citenamefont {Chuang}, \citenamefont {Li},
  \citenamefont {Liu},\ and\ \citenamefont {Lu}}]{Laroche2016}%
  \BibitemOpen
  \bibfield  {author} {\bibinfo {author} {\bibfnamefont {D.}~\bibnamefont
  {Laroche}}, \bibinfo {author} {\bibfnamefont {S.~H.}\ \bibnamefont {Huang}},
  \bibinfo {author} {\bibfnamefont {Y.}~\bibnamefont {Chuang}}, \bibinfo
  {author} {\bibfnamefont {J.~Y.}\ \bibnamefont {Li}}, \bibinfo {author}
  {\bibfnamefont {C.~W.}\ \bibnamefont {Liu}}, \ and\ \bibinfo {author}
  {\bibfnamefont {T.~M.}\ \bibnamefont {Lu}},\ }\bibfield  {title} {\enquote
  {\bibinfo {title} {{Magneto-transport analysis of an ultra-low-density
  two-dimensional hole gas in an undoped strained Ge/SiGe heterostructure}},}\
  }\href {\doibase 10.1063/1.4953399} {\bibfield  {journal} {\bibinfo
  {journal} {Applied Physics Letters}\ }\textbf {\bibinfo {volume} {108}}
  (\bibinfo {year} {2016}),\ 10.1063/1.4953399}\BibitemShut {NoStop}%
\bibitem [{\citenamefont {Sammak}\ \emph {et~al.}(2019)\citenamefont {Sammak},
  \citenamefont {Sabbagh}, \citenamefont {Hendrickx}, \citenamefont {Lodari},
  \citenamefont {{Paquelet Wuetz}}, \citenamefont {Tosato}, \citenamefont
  {Yeoh}, \citenamefont {Bollani}, \citenamefont {Virgilio}, \citenamefont
  {Schubert}, \citenamefont {Zaumseil}, \citenamefont {Capellini},
  \citenamefont {Veldhorst},\ and\ \citenamefont {Scappucci}}]{Sammak2019}%
  \BibitemOpen
  \bibfield  {author} {\bibinfo {author} {\bibfnamefont {A.}~\bibnamefont
  {Sammak}}, \bibinfo {author} {\bibfnamefont {D.}~\bibnamefont {Sabbagh}},
  \bibinfo {author} {\bibfnamefont {N.~W.}\ \bibnamefont {Hendrickx}}, \bibinfo
  {author} {\bibfnamefont {M.}~\bibnamefont {Lodari}}, \bibinfo {author}
  {\bibfnamefont {B.}~\bibnamefont {{Paquelet Wuetz}}}, \bibinfo {author}
  {\bibfnamefont {A.}~\bibnamefont {Tosato}}, \bibinfo {author} {\bibfnamefont
  {L.~R.}\ \bibnamefont {Yeoh}}, \bibinfo {author} {\bibfnamefont
  {M.}~\bibnamefont {Bollani}}, \bibinfo {author} {\bibfnamefont
  {M.}~\bibnamefont {Virgilio}}, \bibinfo {author} {\bibfnamefont {M.~A.}\
  \bibnamefont {Schubert}}, \bibinfo {author} {\bibfnamefont {P.}~\bibnamefont
  {Zaumseil}}, \bibinfo {author} {\bibfnamefont {G.}~\bibnamefont {Capellini}},
  \bibinfo {author} {\bibfnamefont {M.}~\bibnamefont {Veldhorst}}, \ and\
  \bibinfo {author} {\bibfnamefont {G.}~\bibnamefont {Scappucci}},\ }\bibfield
  {title} {\enquote {\bibinfo {title} {{Shallow and Undoped Germanium Quantum
  Wells: A Playground for Spin and Hybrid Quantum Technology}},}\ }\href
  {\doibase 10.1002/adfm.201807613} {\bibfield  {journal} {\bibinfo  {journal}
  {Advanced Functional Materials}\ }\textbf {\bibinfo {volume} {29}} (\bibinfo
  {year} {2019}),\ 10.1002/adfm.201807613}\BibitemShut {NoStop}%
\bibitem [{\citenamefont {Hardy}\ \emph {et~al.}(2019)\citenamefont {Hardy},
  \citenamefont {Harris}, \citenamefont {Su}, \citenamefont {Chuang},
  \citenamefont {Moussa}, \citenamefont {Maurer}, \citenamefont {Li},
  \citenamefont {Lu},\ and\ \citenamefont {Luhman}}]{Hardy2019}%
  \BibitemOpen
  \bibfield  {author} {\bibinfo {author} {\bibfnamefont {W.~J.}\ \bibnamefont
  {Hardy}}, \bibinfo {author} {\bibfnamefont {C.~T.}\ \bibnamefont {Harris}},
  \bibinfo {author} {\bibfnamefont {Y.~H.}\ \bibnamefont {Su}}, \bibinfo
  {author} {\bibfnamefont {Y.}~\bibnamefont {Chuang}}, \bibinfo {author}
  {\bibfnamefont {J.}~\bibnamefont {Moussa}}, \bibinfo {author} {\bibfnamefont
  {L.~N.}\ \bibnamefont {Maurer}}, \bibinfo {author} {\bibfnamefont {J.~Y.}\
  \bibnamefont {Li}}, \bibinfo {author} {\bibfnamefont {T.~M.}\ \bibnamefont
  {Lu}}, \ and\ \bibinfo {author} {\bibfnamefont {D.~R.}\ \bibnamefont
  {Luhman}},\ }\bibfield  {title} {\enquote {\bibinfo {title} {{Single and
  double hole quantum dots in strained Ge/SiGe quantum wells}},}\ }\href
  {\doibase 10.1088/1361-6528/ab061e} {\bibfield  {journal} {\bibinfo
  {journal} {Nanotechnology}\ }\textbf {\bibinfo {volume} {30}} (\bibinfo
  {year} {2019}),\ 10.1088/1361-6528/ab061e}\BibitemShut {NoStop}%
\bibitem [{\citenamefont {Hendrickx}\ \emph {et~al.}(2018)\citenamefont
  {Hendrickx}, \citenamefont {Franke}, \citenamefont {Sammak}, \citenamefont
  {Kouwenhoven}, \citenamefont {Sabbagh}, \citenamefont {Yeoh}, \citenamefont
  {Li}, \citenamefont {Tagliaferri}, \citenamefont {Virgilio}, \citenamefont
  {Capellini}, \citenamefont {Scappucci},\ and\ \citenamefont
  {Veldhorst}}]{Hendrickx2018}%
  \BibitemOpen
  \bibfield  {author} {\bibinfo {author} {\bibfnamefont {N.~W.}\ \bibnamefont
  {Hendrickx}}, \bibinfo {author} {\bibfnamefont {D.~P.}\ \bibnamefont
  {Franke}}, \bibinfo {author} {\bibfnamefont {A.}~\bibnamefont {Sammak}},
  \bibinfo {author} {\bibfnamefont {M.}~\bibnamefont {Kouwenhoven}}, \bibinfo
  {author} {\bibfnamefont {D.}~\bibnamefont {Sabbagh}}, \bibinfo {author}
  {\bibfnamefont {L.}~\bibnamefont {Yeoh}}, \bibinfo {author} {\bibfnamefont
  {R.}~\bibnamefont {Li}}, \bibinfo {author} {\bibfnamefont {M.~L.}\
  \bibnamefont {Tagliaferri}}, \bibinfo {author} {\bibfnamefont
  {M.}~\bibnamefont {Virgilio}}, \bibinfo {author} {\bibfnamefont
  {G.}~\bibnamefont {Capellini}}, \bibinfo {author} {\bibfnamefont
  {G.}~\bibnamefont {Scappucci}}, \ and\ \bibinfo {author} {\bibfnamefont
  {M.}~\bibnamefont {Veldhorst}},\ }\bibfield  {title} {\enquote {\bibinfo
  {title} {{Gate-controlled quantum dots and superconductivity in planar
  germanium}},}\ }\href {\doibase 10.1038/s41467-018-05299-x} {\bibfield
  {journal} {\bibinfo  {journal} {Nature Communications}\ }\textbf {\bibinfo
  {volume} {9}},\ \bibinfo {pages} {1--7} (\bibinfo {year} {2018})},\ \Eprint
  {http://arxiv.org/abs/1801.08869} {arXiv:1801.08869} \BibitemShut {NoStop}%
\bibitem [{\citenamefont {Hendrickx}\ \emph
  {et~al.}(2020{\natexlab{a}})\citenamefont {Hendrickx}, \citenamefont
  {Lawrie}, \citenamefont {Petit}, \citenamefont {Sammak}, \citenamefont
  {Scappucci},\ and\ \citenamefont {Veldhorst}}]{Hendrickx2020}%
  \BibitemOpen
  \bibfield  {author} {\bibinfo {author} {\bibfnamefont {N.~W.}\ \bibnamefont
  {Hendrickx}}, \bibinfo {author} {\bibfnamefont {W.~I.~L.}\ \bibnamefont
  {Lawrie}}, \bibinfo {author} {\bibfnamefont {L.}~\bibnamefont {Petit}},
  \bibinfo {author} {\bibfnamefont {A.}~\bibnamefont {Sammak}}, \bibinfo
  {author} {\bibfnamefont {G.}~\bibnamefont {Scappucci}}, \ and\ \bibinfo
  {author} {\bibfnamefont {M.}~\bibnamefont {Veldhorst}},\ }\bibfield  {title}
  {\enquote {\bibinfo {title} {{A single-hole spin qubit}},}\ }\href {\doibase
  10.1038/s41467-020-17211-7} {\bibfield  {journal} {\bibinfo  {journal}
  {Nature Communications}\ }\textbf {\bibinfo {volume} {11}} (\bibinfo {year}
  {2020}{\natexlab{a}}),\ 10.1038/s41467-020-17211-7},\ \Eprint
  {http://arxiv.org/abs/1912.10426} {arXiv:1912.10426} \BibitemShut {NoStop}%
\bibitem [{\citenamefont {Hendrickx}\ \emph {et~al.}(2021)\citenamefont
  {Hendrickx}, \citenamefont {Lawrie}, \citenamefont {Russ}, \citenamefont {van
  Riggelen}, \citenamefont {de~Snoo}, \citenamefont {Schouten}, \citenamefont
  {Sammak}, \citenamefont {Scappucci},\ and\ \citenamefont
  {Veldhorst}}]{Hendrickx2021}%
  \BibitemOpen
  \bibfield  {author} {\bibinfo {author} {\bibfnamefont {N.~W.}\ \bibnamefont
  {Hendrickx}}, \bibinfo {author} {\bibfnamefont {W.~I.}\ \bibnamefont
  {Lawrie}}, \bibinfo {author} {\bibfnamefont {M.}~\bibnamefont {Russ}},
  \bibinfo {author} {\bibfnamefont {F.}~\bibnamefont {van Riggelen}}, \bibinfo
  {author} {\bibfnamefont {S.~L.}\ \bibnamefont {de~Snoo}}, \bibinfo {author}
  {\bibfnamefont {R.~N.}\ \bibnamefont {Schouten}}, \bibinfo {author}
  {\bibfnamefont {A.}~\bibnamefont {Sammak}}, \bibinfo {author} {\bibfnamefont
  {G.}~\bibnamefont {Scappucci}}, \ and\ \bibinfo {author} {\bibfnamefont
  {M.}~\bibnamefont {Veldhorst}},\ }\bibfield  {title} {\enquote {\bibinfo
  {title} {{A four-qubit germanium quantum processor}},}\ }\href {\doibase
  10.1038/s41586-021-03332-6} {\bibfield  {journal} {\bibinfo  {journal}
  {Nature}\ }\textbf {\bibinfo {volume} {591}},\ \bibinfo {pages} {580--585}
  (\bibinfo {year} {2021})},\ \Eprint {http://arxiv.org/abs/2009.04268}
  {arXiv:2009.04268} \BibitemShut {NoStop}%
\bibitem [{\citenamefont {Hendrickx}\ \emph
  {et~al.}(2020{\natexlab{b}})\citenamefont {Hendrickx}, \citenamefont
  {Franke}, \citenamefont {Sammak}, \citenamefont {Scappucci},\ and\
  \citenamefont {Veldhorst}}]{Hendrickx2020a}%
  \BibitemOpen
  \bibfield  {author} {\bibinfo {author} {\bibfnamefont {N.~W.}\ \bibnamefont
  {Hendrickx}}, \bibinfo {author} {\bibfnamefont {D.~P.}\ \bibnamefont
  {Franke}}, \bibinfo {author} {\bibfnamefont {A.}~\bibnamefont {Sammak}},
  \bibinfo {author} {\bibfnamefont {G.}~\bibnamefont {Scappucci}}, \ and\
  \bibinfo {author} {\bibfnamefont {M.}~\bibnamefont {Veldhorst}},\ }\bibfield
  {title} {\enquote {\bibinfo {title} {{Fast two-qubit logic with holes in
  germanium}},}\ }\href {\doibase 10.1038/s41586-019-1919-3} {\bibfield
  {journal} {\bibinfo  {journal} {Nature}\ }\textbf {\bibinfo {volume} {577}},\
  \bibinfo {pages} {487--491} (\bibinfo {year} {2020}{\natexlab{b}})},\ \Eprint
  {http://arxiv.org/abs/1904.11443} {arXiv:1904.11443} \BibitemShut {NoStop}%
\bibitem [{\citenamefont {Miller}\ \emph {et~al.}(2021)\citenamefont {Miller},
  \citenamefont {Brickson}, \citenamefont {Hardy}, \citenamefont {Liu},
  \citenamefont {Li}, \citenamefont {Baczewski}, \citenamefont {Lilly},
  \citenamefont {Lu},\ and\ \citenamefont {Luhman}}]{Miller2021}%
  \BibitemOpen
  \bibfield  {author} {\bibinfo {author} {\bibfnamefont {A.~J.}\ \bibnamefont
  {Miller}}, \bibinfo {author} {\bibfnamefont {M.}~\bibnamefont {Brickson}},
  \bibinfo {author} {\bibfnamefont {W.~J.}\ \bibnamefont {Hardy}}, \bibinfo
  {author} {\bibfnamefont {C.-Y.}\ \bibnamefont {Liu}}, \bibinfo {author}
  {\bibfnamefont {J.-Y.}\ \bibnamefont {Li}}, \bibinfo {author} {\bibfnamefont
  {A.}~\bibnamefont {Baczewski}}, \bibinfo {author} {\bibfnamefont {M.~P.}\
  \bibnamefont {Lilly}}, \bibinfo {author} {\bibfnamefont {T.-M.}\ \bibnamefont
  {Lu}}, \ and\ \bibinfo {author} {\bibfnamefont {D.~R.}\ \bibnamefont
  {Luhman}},\ }\bibfield  {title} {\enquote {\bibinfo {title} {{Effective
  out-of-plane g-factor in strained-Ge/SiGe quantum dots}},}\ }\href
  {http://arxiv.org/abs/2102.01758} {\  (\bibinfo {year} {2021})},\ \Eprint
  {http://arxiv.org/abs/2102.01758} {arXiv:2102.01758} \BibitemShut {NoStop}%
\bibitem [{\citenamefont {Abadillo-Uriel}\ and\ \citenamefont
  {Calder{\'{o}}n}(2016)}]{Abadillo-Uriel2016}%
  \BibitemOpen
  \bibfield  {author} {\bibinfo {author} {\bibfnamefont {J.~C.}\ \bibnamefont
  {Abadillo-Uriel}}\ and\ \bibinfo {author} {\bibfnamefont {M.~J.}\
  \bibnamefont {Calder{\'{o}}n}},\ }\bibfield  {title} {\enquote {\bibinfo
  {title} {{Interface effects on acceptor qubits in silicon and germanium}},}\
  }\href {\doibase 10.1088/0957-4484/27/2/024003} {\bibfield  {journal}
  {\bibinfo  {journal} {Nanotechnology}\ }\textbf {\bibinfo {volume} {27}}
  (\bibinfo {year} {2016}),\ 10.1088/0957-4484/27/2/024003}\BibitemShut
  {NoStop}%
\bibitem [{\citenamefont {Kobayashi}\ \emph {et~al.}(2021)\citenamefont
  {Kobayashi}, \citenamefont {Salfi}, \citenamefont {Chua}, \citenamefont
  {van~der Heijden}, \citenamefont {House}, \citenamefont {Culcer},
  \citenamefont {Hutchison}, \citenamefont {Johnson}, \citenamefont {McCallum},
  \citenamefont {Riemann}, \citenamefont {Abrosimov}, \citenamefont {Becker},
  \citenamefont {Pohl}, \citenamefont {Simmons},\ and\ \citenamefont
  {Rogge}}]{Kobayashi2021}%
  \BibitemOpen
  \bibfield  {author} {\bibinfo {author} {\bibfnamefont {T.}~\bibnamefont
  {Kobayashi}}, \bibinfo {author} {\bibfnamefont {J.}~\bibnamefont {Salfi}},
  \bibinfo {author} {\bibfnamefont {C.}~\bibnamefont {Chua}}, \bibinfo {author}
  {\bibfnamefont {J.}~\bibnamefont {van~der Heijden}}, \bibinfo {author}
  {\bibfnamefont {M.~G.}\ \bibnamefont {House}}, \bibinfo {author}
  {\bibfnamefont {D.}~\bibnamefont {Culcer}}, \bibinfo {author} {\bibfnamefont
  {W.~D.}\ \bibnamefont {Hutchison}}, \bibinfo {author} {\bibfnamefont {B.~C.}\
  \bibnamefont {Johnson}}, \bibinfo {author} {\bibfnamefont {J.~C.}\
  \bibnamefont {McCallum}}, \bibinfo {author} {\bibfnamefont {H.}~\bibnamefont
  {Riemann}}, \bibinfo {author} {\bibfnamefont {N.~V.}\ \bibnamefont
  {Abrosimov}}, \bibinfo {author} {\bibfnamefont {P.}~\bibnamefont {Becker}},
  \bibinfo {author} {\bibfnamefont {H.~J.}\ \bibnamefont {Pohl}}, \bibinfo
  {author} {\bibfnamefont {M.~Y.}\ \bibnamefont {Simmons}}, \ and\ \bibinfo
  {author} {\bibfnamefont {S.}~\bibnamefont {Rogge}},\ }\bibfield  {title}
  {\enquote {\bibinfo {title} {{Engineering long spin coherence times of
  spin–orbit qubits in silicon}},}\ }\href {\doibase
  10.1038/s41563-020-0743-3} {\bibfield  {journal} {\bibinfo  {journal} {Nature
  Materials}\ }\textbf {\bibinfo {volume} {20}},\ \bibinfo {pages} {38--42}
  (\bibinfo {year} {2021})}\BibitemShut {NoStop}%
\bibitem [{\citenamefont {Asaad}\ \emph {et~al.}(2020)\citenamefont {Asaad},
  \citenamefont {Mourik}, \citenamefont {Joecker}, \citenamefont {Johnson},
  \citenamefont {Baczewski}, \citenamefont {Firgau}, \citenamefont
  {M{\c{a}}dzik}, \citenamefont {Schmitt}, \citenamefont {Pla}, \citenamefont
  {Hudson}, \citenamefont {Itoh}, \citenamefont {McCallum}, \citenamefont
  {Dzurak}, \citenamefont {Laucht},\ and\ \citenamefont {Morello}}]{Asaad2020}%
  \BibitemOpen
  \bibfield  {author} {\bibinfo {author} {\bibfnamefont {S.}~\bibnamefont
  {Asaad}}, \bibinfo {author} {\bibfnamefont {V.}~\bibnamefont {Mourik}},
  \bibinfo {author} {\bibfnamefont {B.}~\bibnamefont {Joecker}}, \bibinfo
  {author} {\bibfnamefont {M.~A.}\ \bibnamefont {Johnson}}, \bibinfo {author}
  {\bibfnamefont {A.~D.}\ \bibnamefont {Baczewski}}, \bibinfo {author}
  {\bibfnamefont {H.~R.}\ \bibnamefont {Firgau}}, \bibinfo {author}
  {\bibfnamefont {M.~T.}\ \bibnamefont {M{\c{a}}dzik}}, \bibinfo {author}
  {\bibfnamefont {V.}~\bibnamefont {Schmitt}}, \bibinfo {author} {\bibfnamefont
  {J.~J.}\ \bibnamefont {Pla}}, \bibinfo {author} {\bibfnamefont {F.~E.}\
  \bibnamefont {Hudson}}, \bibinfo {author} {\bibfnamefont {K.~M.}\
  \bibnamefont {Itoh}}, \bibinfo {author} {\bibfnamefont {J.~C.}\ \bibnamefont
  {McCallum}}, \bibinfo {author} {\bibfnamefont {A.~S.}\ \bibnamefont
  {Dzurak}}, \bibinfo {author} {\bibfnamefont {A.}~\bibnamefont {Laucht}}, \
  and\ \bibinfo {author} {\bibfnamefont {A.}~\bibnamefont {Morello}},\
  }\bibfield  {title} {\enquote {\bibinfo {title} {{Coherent electrical control
  of a single high-spin nucleus in silicon}},}\ }\href {\doibase
  10.1038/s41586-020-2057-7} {\bibfield  {journal} {\bibinfo  {journal}
  {Nature}\ }\textbf {\bibinfo {volume} {579}},\ \bibinfo {pages} {205--209}
  (\bibinfo {year} {2020})},\ \Eprint {http://arxiv.org/abs/1906.01086}
  {arXiv:1906.01086} \BibitemShut {NoStop}%
\bibitem [{\citenamefont {Satta}\ \emph {et~al.}(2006)\citenamefont {Satta},
  \citenamefont {Simoen}, \citenamefont {Janssens}, \citenamefont {Clarysse},
  \citenamefont {{De Jaeger}}, \citenamefont {Benedetti}, \citenamefont
  {Hoflijk}, \citenamefont {Brijs}, \citenamefont {Meuris},\ and\ \citenamefont
  {Vandervorst}}]{Satta2006}%
  \BibitemOpen
  \bibfield  {author} {\bibinfo {author} {\bibfnamefont {A.}~\bibnamefont
  {Satta}}, \bibinfo {author} {\bibfnamefont {E.}~\bibnamefont {Simoen}},
  \bibinfo {author} {\bibfnamefont {T.}~\bibnamefont {Janssens}}, \bibinfo
  {author} {\bibfnamefont {T.}~\bibnamefont {Clarysse}}, \bibinfo {author}
  {\bibfnamefont {B.}~\bibnamefont {{De Jaeger}}}, \bibinfo {author}
  {\bibfnamefont {A.}~\bibnamefont {Benedetti}}, \bibinfo {author}
  {\bibfnamefont {I.}~\bibnamefont {Hoflijk}}, \bibinfo {author} {\bibfnamefont
  {B.}~\bibnamefont {Brijs}}, \bibinfo {author} {\bibfnamefont
  {M.}~\bibnamefont {Meuris}}, \ and\ \bibinfo {author} {\bibfnamefont
  {W.}~\bibnamefont {Vandervorst}},\ }\bibfield  {title} {\enquote {\bibinfo
  {title} {{Shallow Junction Ion Implantation in Ge and Associated Defect
  Control}},}\ }\href {\doibase 10.1149/1.2162469} {\bibfield  {journal}
  {\bibinfo  {journal} {Journal of The Electrochemical Society}\ }\textbf
  {\bibinfo {volume} {153}},\ \bibinfo {pages} {G229} (\bibinfo {year}
  {2006})}\BibitemShut {NoStop}%
\bibitem [{\citenamefont {Murrell}\ \emph {et~al.}(2000)\citenamefont
  {Murrell}, \citenamefont {Collart}, \citenamefont {Foad},\ and\ \citenamefont
  {Jennings}}]{Murrell2000}%
  \BibitemOpen
  \bibfield  {author} {\bibinfo {author} {\bibfnamefont {A.~J.}\ \bibnamefont
  {Murrell}}, \bibinfo {author} {\bibfnamefont {E.~J.~H.}\ \bibnamefont
  {Collart}}, \bibinfo {author} {\bibfnamefont {M.~A.}\ \bibnamefont {Foad}}, \
  and\ \bibinfo {author} {\bibfnamefont {D.}~\bibnamefont {Jennings}},\
  }\bibfield  {title} {\enquote {\bibinfo {title} {{Process interactions
  between low-energy ion implantation and rapid-thermal annealing for optimized
  ultrashallow junction formation}},}\ }\href {\doibase 10.1116/1.591212}
  {\bibfield  {journal} {\bibinfo  {journal} {Journal of Vacuum Science {\&}
  Technology B: Microelectronics and Nanometer Structures}\ }\textbf {\bibinfo
  {volume} {18}},\ \bibinfo {pages} {462} (\bibinfo {year} {2000})}\BibitemShut
  {NoStop}%
\bibitem [{\citenamefont {Impellizzeri}\ \emph {et~al.}(2009)\citenamefont
  {Impellizzeri}, \citenamefont {Mirabella}, \citenamefont {Irrera},
  \citenamefont {Grimaldi},\ and\ \citenamefont
  {Napolitani}}]{Impellizzeri2009}%
  \BibitemOpen
  \bibfield  {author} {\bibinfo {author} {\bibfnamefont {G.}~\bibnamefont
  {Impellizzeri}}, \bibinfo {author} {\bibfnamefont {S.}~\bibnamefont
  {Mirabella}}, \bibinfo {author} {\bibfnamefont {A.}~\bibnamefont {Irrera}},
  \bibinfo {author} {\bibfnamefont {M.~G.}\ \bibnamefont {Grimaldi}}, \ and\
  \bibinfo {author} {\bibfnamefont {E.}~\bibnamefont {Napolitani}},\ }\bibfield
   {title} {\enquote {\bibinfo {title} {{Ga-implantation in Ge: Electrical
  activation and clustering}},}\ }\href {\doibase 10.1063/1.3159031} {\bibfield
   {journal} {\bibinfo  {journal} {Journal of Applied Physics}\ }\textbf
  {\bibinfo {volume} {106}} (\bibinfo {year} {2009}),\
  10.1063/1.3159031}\BibitemShut {NoStop}%
\bibitem [{\citenamefont {Heera}\ \emph {et~al.}(2010)\citenamefont {Heera},
  \citenamefont {M{\"{u}}cklich}, \citenamefont {Posselt}, \citenamefont
  {Voelskow}, \citenamefont {W{\"{u}}ndisch}, \citenamefont {Schmidt},
  \citenamefont {Skrotzki}, \citenamefont {Heinig}, \citenamefont
  {Herrmannsd{\"{o}}rfer},\ and\ \citenamefont {Skorupa}}]{Heera2010}%
  \BibitemOpen
  \bibfield  {author} {\bibinfo {author} {\bibfnamefont {V.}~\bibnamefont
  {Heera}}, \bibinfo {author} {\bibfnamefont {A.}~\bibnamefont
  {M{\"{u}}cklich}}, \bibinfo {author} {\bibfnamefont {M.}~\bibnamefont
  {Posselt}}, \bibinfo {author} {\bibfnamefont {M.}~\bibnamefont {Voelskow}},
  \bibinfo {author} {\bibfnamefont {C.}~\bibnamefont {W{\"{u}}ndisch}},
  \bibinfo {author} {\bibfnamefont {B.}~\bibnamefont {Schmidt}}, \bibinfo
  {author} {\bibfnamefont {R.}~\bibnamefont {Skrotzki}}, \bibinfo {author}
  {\bibfnamefont {K.~H.}\ \bibnamefont {Heinig}}, \bibinfo {author}
  {\bibfnamefont {T.}~\bibnamefont {Herrmannsd{\"{o}}rfer}}, \ and\ \bibinfo
  {author} {\bibfnamefont {W.}~\bibnamefont {Skorupa}},\ }\bibfield  {title}
  {\enquote {\bibinfo {title} {{Heavily Ga-doped germanium layers produced by
  ion implantation and flash lamp annealing: Structure and electrical
  activation}},}\ }\href {\doibase 10.1063/1.3309835} {\bibfield  {journal}
  {\bibinfo  {journal} {Journal of Applied Physics}\ }\textbf {\bibinfo
  {volume} {107}},\ \bibinfo {pages} {053508} (\bibinfo {year}
  {2010})}\BibitemShut {NoStop}%
\bibitem [{\citenamefont {Hellings}\ \emph {et~al.}(2009)\citenamefont
  {Hellings}, \citenamefont {Wuendisch}, \citenamefont {Eneman}, \citenamefont
  {Simoen}, \citenamefont {Clarysse}, \citenamefont {Meuris}, \citenamefont
  {Vandervorst}, \citenamefont {Posselt},\ and\ \citenamefont {{De
  Meyer}}}]{Hellings2009}%
  \BibitemOpen
  \bibfield  {author} {\bibinfo {author} {\bibfnamefont {G.}~\bibnamefont
  {Hellings}}, \bibinfo {author} {\bibfnamefont {C.}~\bibnamefont {Wuendisch}},
  \bibinfo {author} {\bibfnamefont {G.}~\bibnamefont {Eneman}}, \bibinfo
  {author} {\bibfnamefont {E.}~\bibnamefont {Simoen}}, \bibinfo {author}
  {\bibfnamefont {T.}~\bibnamefont {Clarysse}}, \bibinfo {author}
  {\bibfnamefont {M.}~\bibnamefont {Meuris}}, \bibinfo {author} {\bibfnamefont
  {W.}~\bibnamefont {Vandervorst}}, \bibinfo {author} {\bibfnamefont
  {M.}~\bibnamefont {Posselt}}, \ and\ \bibinfo {author} {\bibfnamefont
  {K.}~\bibnamefont {{De Meyer}}},\ }\bibfield  {title} {\enquote {\bibinfo
  {title} {{Implantation, diffusion, activation, and recrystallization of
  gallium implanted in preamorphized and crystalline germanium}},}\ }\href
  {\doibase 10.1149/1.3225204} {\bibfield  {journal} {\bibinfo  {journal}
  {Electrochemical and Solid-State Letters}\ }\textbf {\bibinfo {volume}
  {12}},\ \bibinfo {pages} {417--420} (\bibinfo {year} {2009})}\BibitemShut
  {NoStop}%
\bibitem [{\citenamefont {Morello}\ \emph {et~al.}(2010)\citenamefont
  {Morello}, \citenamefont {Pla}, \citenamefont {Zwanenburg}, \citenamefont
  {Chan}, \citenamefont {Tan}, \citenamefont {Huebl}, \citenamefont
  {M{\"{o}}tt{\"{o}}nen}, \citenamefont {Nugroho}, \citenamefont {Yang},
  \citenamefont {{Van Donkelaar}}, \citenamefont {Alves}, \citenamefont
  {Jamieson}, \citenamefont {Escott}, \citenamefont {Hollenberg}, \citenamefont
  {Clark},\ and\ \citenamefont {Dzurak}}]{Morello2010}%
  \BibitemOpen
  \bibfield  {author} {\bibinfo {author} {\bibfnamefont {A.}~\bibnamefont
  {Morello}}, \bibinfo {author} {\bibfnamefont {J.~J.}\ \bibnamefont {Pla}},
  \bibinfo {author} {\bibfnamefont {F.~A.}\ \bibnamefont {Zwanenburg}},
  \bibinfo {author} {\bibfnamefont {K.~W.}\ \bibnamefont {Chan}}, \bibinfo
  {author} {\bibfnamefont {K.~Y.}\ \bibnamefont {Tan}}, \bibinfo {author}
  {\bibfnamefont {H.}~\bibnamefont {Huebl}}, \bibinfo {author} {\bibfnamefont
  {M.}~\bibnamefont {M{\"{o}}tt{\"{o}}nen}}, \bibinfo {author} {\bibfnamefont
  {C.~D.}\ \bibnamefont {Nugroho}}, \bibinfo {author} {\bibfnamefont
  {C.}~\bibnamefont {Yang}}, \bibinfo {author} {\bibfnamefont {J.~A.}\
  \bibnamefont {{Van Donkelaar}}}, \bibinfo {author} {\bibfnamefont {A.~D.}\
  \bibnamefont {Alves}}, \bibinfo {author} {\bibfnamefont {D.~N.}\ \bibnamefont
  {Jamieson}}, \bibinfo {author} {\bibfnamefont {C.~C.}\ \bibnamefont
  {Escott}}, \bibinfo {author} {\bibfnamefont {L.~C.}\ \bibnamefont
  {Hollenberg}}, \bibinfo {author} {\bibfnamefont {R.~G.}\ \bibnamefont
  {Clark}}, \ and\ \bibinfo {author} {\bibfnamefont {A.~S.}\ \bibnamefont
  {Dzurak}},\ }\bibfield  {title} {\enquote {\bibinfo {title} {{Single-shot
  readout of an electron spin in silicon}},}\ }\href {\doibase
  10.1038/nature09392} {\bibfield  {journal} {\bibinfo  {journal} {Nature}\
  }\textbf {\bibinfo {volume} {467}},\ \bibinfo {pages} {687--691} (\bibinfo
  {year} {2010})},\ \Eprint {http://arxiv.org/abs/1003.2679} {arXiv:1003.2679}
  \BibitemShut {NoStop}%
\bibitem [{\citenamefont {Tracy}\ \emph {et~al.}(2013)\citenamefont {Tracy},
  \citenamefont {Lu}, \citenamefont {Bishop}, \citenamefont {{Ten Eyck}},
  \citenamefont {Pluym}, \citenamefont {Wendt}, \citenamefont {Lilly},\ and\
  \citenamefont {Carroll}}]{Tracy2013}%
  \BibitemOpen
  \bibfield  {author} {\bibinfo {author} {\bibfnamefont {L.~A.}\ \bibnamefont
  {Tracy}}, \bibinfo {author} {\bibfnamefont {T.~M.}\ \bibnamefont {Lu}},
  \bibinfo {author} {\bibfnamefont {N.~C.}\ \bibnamefont {Bishop}}, \bibinfo
  {author} {\bibfnamefont {G.~A.}\ \bibnamefont {{Ten Eyck}}}, \bibinfo
  {author} {\bibfnamefont {T.}~\bibnamefont {Pluym}}, \bibinfo {author}
  {\bibfnamefont {J.~R.}\ \bibnamefont {Wendt}}, \bibinfo {author}
  {\bibfnamefont {M.~P.}\ \bibnamefont {Lilly}}, \ and\ \bibinfo {author}
  {\bibfnamefont {M.~S.}\ \bibnamefont {Carroll}},\ }\bibfield  {title}
  {\enquote {\bibinfo {title} {{Electron spin lifetime of a single antimony
  donor in silicon}},}\ }\href {\doibase 10.1063/1.4824128} {\bibfield
  {journal} {\bibinfo  {journal} {Applied Physics Letters}\ }\textbf {\bibinfo
  {volume} {103}},\ \bibinfo {pages} {2--6} (\bibinfo {year}
  {2013})}\BibitemShut {NoStop}%
\bibitem [{\citenamefont {Ziegler}, \citenamefont {Ziegler},\ and\
  \citenamefont {Biersack}(2010)}]{Ziegler2010}%
  \BibitemOpen
  \bibfield  {author} {\bibinfo {author} {\bibfnamefont {J.~F.}\ \bibnamefont
  {Ziegler}}, \bibinfo {author} {\bibfnamefont {M.~D.}\ \bibnamefont
  {Ziegler}}, \ and\ \bibinfo {author} {\bibfnamefont {J.~P.}\ \bibnamefont
  {Biersack}},\ }\bibfield  {title} {\enquote {\bibinfo {title} {{SRIM - The
  stopping and range of ions in matter (2010)}},}\ }\href {\doibase
  10.1016/j.nimb.2010.02.091} {\bibfield  {journal} {\bibinfo  {journal}
  {Nuclear Instruments and Methods in Physics Research, Section B: Beam
  Interactions with Materials and Atoms}\ }\textbf {\bibinfo {volume} {268}},\
  \bibinfo {pages} {1818--1823} (\bibinfo {year} {2010})}\BibitemShut {NoStop}%
\bibitem [{\citenamefont {Sumikura}\ \emph {et~al.}(2011)\citenamefont
  {Sumikura}, \citenamefont {Nishiguchi}, \citenamefont {Ono}, \citenamefont
  {Fujiwara},\ and\ \citenamefont {Notomi}}]{Sumikura2011}%
  \BibitemOpen
  \bibfield  {author} {\bibinfo {author} {\bibfnamefont {H.}~\bibnamefont
  {Sumikura}}, \bibinfo {author} {\bibfnamefont {K.}~\bibnamefont
  {Nishiguchi}}, \bibinfo {author} {\bibfnamefont {Y.}~\bibnamefont {Ono}},
  \bibinfo {author} {\bibfnamefont {A.}~\bibnamefont {Fujiwara}}, \ and\
  \bibinfo {author} {\bibfnamefont {M.}~\bibnamefont {Notomi}},\ }\bibfield
  {title} {\enquote {\bibinfo {title} {{Bound exciton photoluminescence from
  ion-implanted phosphorus in thin silicon layers}},}\ }\href {\doibase
  10.1364/oe.19.025255} {\bibfield  {journal} {\bibinfo  {journal} {Optics
  Express}\ }\textbf {\bibinfo {volume} {19}},\ \bibinfo {pages} {25255}
  (\bibinfo {year} {2011})}\BibitemShut {NoStop}%
\bibitem [{\citenamefont {Poon}\ \emph {et~al.}(2005)\citenamefont {Poon},
  \citenamefont {Tan}, \citenamefont {Cho},\ and\ \citenamefont
  {Du}}]{Poon2005}%
  \BibitemOpen
  \bibfield  {author} {\bibinfo {author} {\bibfnamefont {C.~H.}\ \bibnamefont
  {Poon}}, \bibinfo {author} {\bibfnamefont {L.~S.}\ \bibnamefont {Tan}},
  \bibinfo {author} {\bibfnamefont {B.~J.}\ \bibnamefont {Cho}}, \ and\
  \bibinfo {author} {\bibfnamefont {A.~Y.}\ \bibnamefont {Du}},\ }\bibfield
  {title} {\enquote {\bibinfo {title} {{Dopant Loss Mechanism in n+p Germanium
  Junctions during Rapid Thermal Annealing}},}\ }\href {\doibase
  10.1149/1.2073048} {\bibfield  {journal} {\bibinfo  {journal} {Journal of The
  Electrochemical Society}\ }\textbf {\bibinfo {volume} {152}},\ \bibinfo
  {pages} {G895} (\bibinfo {year} {2005})}\BibitemShut {NoStop}%
\bibitem [{\citenamefont {Ioannou}\ \emph {et~al.}(2008)\citenamefont
  {Ioannou}, \citenamefont {Skarlatos}, \citenamefont {Tsamis}, \citenamefont
  {Krontiras}, \citenamefont {Georga}, \citenamefont {Christofi},\ and\
  \citenamefont {McPhail}}]{Ioannou2008}%
  \BibitemOpen
  \bibfield  {author} {\bibinfo {author} {\bibfnamefont {N.}~\bibnamefont
  {Ioannou}}, \bibinfo {author} {\bibfnamefont {D.}~\bibnamefont {Skarlatos}},
  \bibinfo {author} {\bibfnamefont {C.}~\bibnamefont {Tsamis}}, \bibinfo
  {author} {\bibfnamefont {C.~A.}\ \bibnamefont {Krontiras}}, \bibinfo {author}
  {\bibfnamefont {S.~N.}\ \bibnamefont {Georga}}, \bibinfo {author}
  {\bibfnamefont {A.}~\bibnamefont {Christofi}}, \ and\ \bibinfo {author}
  {\bibfnamefont {D.~S.}\ \bibnamefont {McPhail}},\ }\bibfield  {title}
  {\enquote {\bibinfo {title} {{Germanium substrate loss during low temperature
  annealing and its influence on ion-implanted phosphorous dose loss}},}\
  }\href {\doibase 10.1063/1.2981522} {\bibfield  {journal} {\bibinfo
  {journal} {Applied Physics Letters}\ }\textbf {\bibinfo {volume} {93}},\
  \bibinfo {pages} {10--13} (\bibinfo {year} {2008})}\BibitemShut {NoStop}%
\bibitem [{\citenamefont {Chroneos}\ and\ \citenamefont
  {Bracht}(2014)}]{Chroneos2014}%
  \BibitemOpen
  \bibfield  {author} {\bibinfo {author} {\bibfnamefont {A.}~\bibnamefont
  {Chroneos}}\ and\ \bibinfo {author} {\bibfnamefont {H.}~\bibnamefont
  {Bracht}},\ }\bibfield  {title} {\enquote {\bibinfo {title} {{Diffusion of n
  -type dopants in germanium}},}\ }\href {\doibase 10.1063/1.4838215}
  {\bibfield  {journal} {\bibinfo  {journal} {Applied Physics Reviews}\
  }\textbf {\bibinfo {volume} {1}} (\bibinfo {year} {2014}),\
  10.1063/1.4838215}\BibitemShut {NoStop}%
\bibitem [{\citenamefont {Lazarenkova}\ and\ \citenamefont
  {Balandin}(2003)}]{Lazarenkova2003}%
  \BibitemOpen
  \bibfield  {author} {\bibinfo {author} {\bibfnamefont {O.~L.}\ \bibnamefont
  {Lazarenkova}}\ and\ \bibinfo {author} {\bibfnamefont {A.~A.}\ \bibnamefont
  {Balandin}},\ }\bibfield  {title} {\enquote {\bibinfo {title} {{Raman
  scattering from three-dimensionally regimented quantum dot superlattices}},}\
  }\href {\doibase 10.1016/S0749-6036(03)00049-1} {\bibfield  {journal}
  {\bibinfo  {journal} {Superlattices and Microstructures}\ }\textbf {\bibinfo
  {volume} {33}},\ \bibinfo {pages} {95--101} (\bibinfo {year}
  {2003})}\BibitemShut {NoStop}%
\bibitem [{\citenamefont {Pizani}\ \emph {et~al.}(2000)\citenamefont {Pizani},
  \citenamefont {Lanciotti}, \citenamefont {Jasinevicius}, \citenamefont
  {Duduch},\ and\ \citenamefont {Porto}}]{Pizani2000}%
  \BibitemOpen
  \bibfield  {author} {\bibinfo {author} {\bibfnamefont {P.~S.}\ \bibnamefont
  {Pizani}}, \bibinfo {author} {\bibfnamefont {F.}~\bibnamefont {Lanciotti}},
  \bibinfo {author} {\bibfnamefont {R.~G.}\ \bibnamefont {Jasinevicius}},
  \bibinfo {author} {\bibfnamefont {J.~G.}\ \bibnamefont {Duduch}}, \ and\
  \bibinfo {author} {\bibfnamefont {A.~J.}\ \bibnamefont {Porto}},\ }\bibfield
  {title} {\enquote {\bibinfo {title} {{Raman characterization of structural
  disorder and residual strains in micromachined GaAs}},}\ }\href {\doibase
  10.1063/1.372009} {\bibfield  {journal} {\bibinfo  {journal} {Journal of
  Applied Physics}\ }\textbf {\bibinfo {volume} {87}},\ \bibinfo {pages}
  {1280--1283} (\bibinfo {year} {2000})}\BibitemShut {NoStop}%
\bibitem [{\citenamefont {Cheng}\ \emph {et~al.}(2013)\citenamefont {Cheng},
  \citenamefont {Wang}, \citenamefont {Gong}, \citenamefont {Sun},
  \citenamefont {Guo}, \citenamefont {Hu}, \citenamefont {Shen}, \citenamefont
  {Han},\ and\ \citenamefont {Yeo}}]{Cheng2013}%
  \BibitemOpen
  \bibfield  {author} {\bibinfo {author} {\bibfnamefont {R.}~\bibnamefont
  {Cheng}}, \bibinfo {author} {\bibfnamefont {W.}~\bibnamefont {Wang}},
  \bibinfo {author} {\bibfnamefont {X.}~\bibnamefont {Gong}}, \bibinfo {author}
  {\bibfnamefont {L.}~\bibnamefont {Sun}}, \bibinfo {author} {\bibfnamefont
  {P.}~\bibnamefont {Guo}}, \bibinfo {author} {\bibfnamefont {H.}~\bibnamefont
  {Hu}}, \bibinfo {author} {\bibfnamefont {Z.}~\bibnamefont {Shen}}, \bibinfo
  {author} {\bibfnamefont {G.}~\bibnamefont {Han}}, \ and\ \bibinfo {author}
  {\bibfnamefont {Y.-C.}\ \bibnamefont {Yeo}},\ }\bibfield  {title} {\enquote
  {\bibinfo {title} {{Relaxed and Strained Patterned Germanium-Tin Structures:
  A Raman Scattering Study}},}\ }\href {\doibase 10.1149/2.013304jss}
  {\bibfield  {journal} {\bibinfo  {journal} {ECS Journal of Solid State
  Science and Technology}\ }\textbf {\bibinfo {volume} {2}},\ \bibinfo {pages}
  {P138--P145} (\bibinfo {year} {2013})}\BibitemShut {NoStop}%
\bibitem [{\citenamefont {Sui}\ and\ \citenamefont {Herman}(1993)}]{Sui1993}%
  \BibitemOpen
  \bibfield  {author} {\bibinfo {author} {\bibfnamefont {Z.}~\bibnamefont
  {Sui}}\ and\ \bibinfo {author} {\bibfnamefont {I.~P.}\ \bibnamefont
  {Herman}},\ }\bibfield  {title} {\enquote {\bibinfo {title} {{Effect of
  strain on phonons in Si, Ge, and SiGe heterostructures}},}\ }\href {\doibase
  10.1103/PhysRevB.48.17938} {\bibfield  {journal} {\bibinfo  {journal}
  {Physical Review B}\ }\textbf {\bibinfo {volume} {48}},\ \bibinfo {pages}
  {17938--17953} (\bibinfo {year} {1993})}\BibitemShut {NoStop}%
\bibitem [{\citenamefont {Yukhymchyk}\ \emph {et~al.}(2015)\citenamefont
  {Yukhymchyk}, \citenamefont {Lytvyn}, \citenamefont {Korchovyi},
  \citenamefont {Okholin}, \citenamefont {Glotov}, \citenamefont {Lysenko},
  \citenamefont {Nazarova}, \citenamefont {Shayesteh}, \citenamefont {Duffy},
  \citenamefont {Nazarov}, \citenamefont {Chemistry}, \citenamefont
  {Maltings},\ and\ \citenamefont {Row}}]{Yukhymchyk2015}%
  \BibitemOpen
  \bibfield  {author} {\bibinfo {author} {\bibfnamefont {V.~O.}\ \bibnamefont
  {Yukhymchyk}}, \bibinfo {author} {\bibfnamefont {P.~M.}\ \bibnamefont
  {Lytvyn}}, \bibinfo {author} {\bibfnamefont {A.~A.}\ \bibnamefont
  {Korchovyi}}, \bibinfo {author} {\bibfnamefont {P.~N.}\ \bibnamefont
  {Okholin}}, \bibinfo {author} {\bibfnamefont {V.~I.}\ \bibnamefont {Glotov}},
  \bibinfo {author} {\bibfnamefont {V.~S.}\ \bibnamefont {Lysenko}}, \bibinfo
  {author} {\bibfnamefont {T.~M.}\ \bibnamefont {Nazarova}}, \bibinfo {author}
  {\bibfnamefont {M.}~\bibnamefont {Shayesteh}}, \bibinfo {author}
  {\bibfnamefont {R.}~\bibnamefont {Duffy}}, \bibinfo {author} {\bibfnamefont
  {A.~N.}\ \bibnamefont {Nazarov}}, \bibinfo {author} {\bibfnamefont
  {N.}~\bibnamefont {Chemistry}}, \bibinfo {author} {\bibfnamefont
  {L.}~\bibnamefont {Maltings}}, \ and\ \bibinfo {author} {\bibfnamefont
  {P.}~\bibnamefont {Row}},\ }\bibfield  {title} {\enquote {\bibinfo {title}
  {{RF plasma effect on amorphous thin ion-implanted layers of n- and p-type
  germanium: Raman and AFM research}},}\ }\href@noop {} {\bibfield  {journal}
  {\bibinfo  {journal} {Proceedings of the 11th International Conference
  "Interaction of Radiation with Solids"}\ ,\ \bibinfo {pages} {23--25}}
  (\bibinfo {year} {2015})}\BibitemShut {NoStop}%
\bibitem [{\citenamefont {Lieten}\ \emph {et~al.}(2013)\citenamefont {Lieten},
  \citenamefont {Douhard}, \citenamefont {Stesmans}, \citenamefont {Jivanescu},
  \citenamefont {Beeman}, \citenamefont {Simoen}, \citenamefont {Vandervorst},
  \citenamefont {Haller},\ and\ \citenamefont {Locquet}}]{Lieten2013}%
  \BibitemOpen
  \bibfield  {author} {\bibinfo {author} {\bibfnamefont {R.~R.}\ \bibnamefont
  {Lieten}}, \bibinfo {author} {\bibfnamefont {B.}~\bibnamefont {Douhard}},
  \bibinfo {author} {\bibfnamefont {A.}~\bibnamefont {Stesmans}}, \bibinfo
  {author} {\bibfnamefont {M.}~\bibnamefont {Jivanescu}}, \bibinfo {author}
  {\bibfnamefont {J.~W.}\ \bibnamefont {Beeman}}, \bibinfo {author}
  {\bibfnamefont {E.}~\bibnamefont {Simoen}}, \bibinfo {author} {\bibfnamefont
  {W.}~\bibnamefont {Vandervorst}}, \bibinfo {author} {\bibfnamefont {E.~E.}\
  \bibnamefont {Haller}}, \ and\ \bibinfo {author} {\bibfnamefont {J.-P.}\
  \bibnamefont {Locquet}},\ }\bibfield  {title} {\enquote {\bibinfo {title}
  {{Implantation and Activation of Phosphorus in Amorphous and Crystalline
  Germanium Layers}},}\ }\href {\doibase 10.1149/2.011309jss} {\bibfield
  {journal} {\bibinfo  {journal} {ECS Journal of Solid State Science and
  Technology}\ }\textbf {\bibinfo {volume} {2}},\ \bibinfo {pages} {P346--P350}
  (\bibinfo {year} {2013})}\BibitemShut {NoStop}%
\bibitem [{\citenamefont {Suh}\ \emph {et~al.}(2005)\citenamefont {Suh},
  \citenamefont {Carroll}, \citenamefont {Levy}, \citenamefont {Bisognin},
  \citenamefont {{De Salvador}},\ and\ \citenamefont {Sahiner}}]{Suh2005}%
  \BibitemOpen
  \bibfield  {author} {\bibinfo {author} {\bibfnamefont {Y.~S.}\ \bibnamefont
  {Suh}}, \bibinfo {author} {\bibfnamefont {M.~S.}\ \bibnamefont {Carroll}},
  \bibinfo {author} {\bibfnamefont {R.~A.}\ \bibnamefont {Levy}}, \bibinfo
  {author} {\bibfnamefont {G.}~\bibnamefont {Bisognin}}, \bibinfo {author}
  {\bibfnamefont {D.}~\bibnamefont {{De Salvador}}}, \ and\ \bibinfo {author}
  {\bibfnamefont {M.~A.}\ \bibnamefont {Sahiner}},\ }\bibfield  {title}
  {\enquote {\bibinfo {title} {{Implantation and activation of high
  concentrations of boron and phosphorus in Germanium}},}\ }\href {\doibase
  10.1557/proc-0891-ee07-20} {\bibfield  {journal} {\bibinfo  {journal} {IEEE
  Transactions of Electron Devices}\ }\textbf {\bibinfo {volume} {52}},\
  \bibinfo {pages} {2416--2421} (\bibinfo {year} {2005})}\BibitemShut {NoStop}%
\bibitem [{\citenamefont {Koffel}\ \emph {et~al.}(2009)\citenamefont {Koffel},
  \citenamefont {Scheiblin}, \citenamefont {Claverie},\ and\ \citenamefont
  {Benassayag}}]{Koffel2009}%
  \BibitemOpen
  \bibfield  {author} {\bibinfo {author} {\bibfnamefont {S.}~\bibnamefont
  {Koffel}}, \bibinfo {author} {\bibfnamefont {P.}~\bibnamefont {Scheiblin}},
  \bibinfo {author} {\bibfnamefont {A.}~\bibnamefont {Claverie}}, \ and\
  \bibinfo {author} {\bibfnamefont {G.}~\bibnamefont {Benassayag}},\ }\bibfield
   {title} {\enquote {\bibinfo {title} {{Amorphization kinetics of germanium
  during ion implantation}},}\ }\href {\doibase 10.1063/1.3041653} {\bibfield
  {journal} {\bibinfo  {journal} {Journal of Applied Physics}\ }\textbf
  {\bibinfo {volume} {105}} (\bibinfo {year} {2009}),\
  10.1063/1.3041653}\BibitemShut {NoStop}%
\bibitem [{\citenamefont {Holmes}\ \emph {et~al.}(2019)\citenamefont {Holmes},
  \citenamefont {Lawrie}, \citenamefont {Johnson}, \citenamefont
  {Asadpoordarvish}, \citenamefont {McCallum}, \citenamefont {McCamey},\ and\
  \citenamefont {Jamieson}}]{Holmes2019}%
  \BibitemOpen
  \bibfield  {author} {\bibinfo {author} {\bibfnamefont {D.}~\bibnamefont
  {Holmes}}, \bibinfo {author} {\bibfnamefont {W.~I.}\ \bibnamefont {Lawrie}},
  \bibinfo {author} {\bibfnamefont {B.~C.}\ \bibnamefont {Johnson}}, \bibinfo
  {author} {\bibfnamefont {A.}~\bibnamefont {Asadpoordarvish}}, \bibinfo
  {author} {\bibfnamefont {J.~C.}\ \bibnamefont {McCallum}}, \bibinfo {author}
  {\bibfnamefont {D.~R.}\ \bibnamefont {McCamey}}, \ and\ \bibinfo {author}
  {\bibfnamefont {D.~N.}\ \bibnamefont {Jamieson}},\ }\bibfield  {title}
  {\enquote {\bibinfo {title} {{Activation and electron spin resonance of
  near-surface implanted bismuth donors in silicon}},}\ }\href {\doibase
  10.1103/PhysRevMaterials.3.083403} {\bibfield  {journal} {\bibinfo  {journal}
  {Physical Review Materials}\ }\textbf {\bibinfo {volume} {3}},\ \bibinfo
  {pages} {1--7} (\bibinfo {year} {2019})}\BibitemShut {NoStop}%
\bibitem [{\citenamefont {Pla}\ \emph {et~al.}(2012)\citenamefont {Pla},
  \citenamefont {Tan}, \citenamefont {Dehollain}, \citenamefont {Lim},
  \citenamefont {Morton}, \citenamefont {Jamieson}, \citenamefont {Dzurak},\
  and\ \citenamefont {Morello}}]{Pla2012}%
  \BibitemOpen
  \bibfield  {author} {\bibinfo {author} {\bibfnamefont {J.~J.}\ \bibnamefont
  {Pla}}, \bibinfo {author} {\bibfnamefont {K.~Y.}\ \bibnamefont {Tan}},
  \bibinfo {author} {\bibfnamefont {J.~P.}\ \bibnamefont {Dehollain}}, \bibinfo
  {author} {\bibfnamefont {W.~H.}\ \bibnamefont {Lim}}, \bibinfo {author}
  {\bibfnamefont {J.~J.}\ \bibnamefont {Morton}}, \bibinfo {author}
  {\bibfnamefont {D.~N.}\ \bibnamefont {Jamieson}}, \bibinfo {author}
  {\bibfnamefont {A.~S.}\ \bibnamefont {Dzurak}}, \ and\ \bibinfo {author}
  {\bibfnamefont {A.}~\bibnamefont {Morello}},\ }\bibfield  {title} {\enquote
  {\bibinfo {title} {{A single-atom electron spin qubit in silicon}},}\ }\href
  {\doibase 10.1038/nature11449} {\bibfield  {journal} {\bibinfo  {journal}
  {Nature}\ }\textbf {\bibinfo {volume} {489}},\ \bibinfo {pages} {541--544}
  (\bibinfo {year} {2012})}\BibitemShut {NoStop}%
\bibitem [{\citenamefont {Pla}\ \emph {et~al.}(2013)\citenamefont {Pla},
  \citenamefont {Tan}, \citenamefont {Dehollain}, \citenamefont {Lim},
  \citenamefont {Morton}, \citenamefont {Zwanenburg}, \citenamefont {Jamieson},
  \citenamefont {Dzurak},\ and\ \citenamefont {Morello}}]{Pla2013}%
  \BibitemOpen
  \bibfield  {author} {\bibinfo {author} {\bibfnamefont {J.~J.}\ \bibnamefont
  {Pla}}, \bibinfo {author} {\bibfnamefont {K.~Y.}\ \bibnamefont {Tan}},
  \bibinfo {author} {\bibfnamefont {J.~P.}\ \bibnamefont {Dehollain}}, \bibinfo
  {author} {\bibfnamefont {W.~H.}\ \bibnamefont {Lim}}, \bibinfo {author}
  {\bibfnamefont {J.~J.}\ \bibnamefont {Morton}}, \bibinfo {author}
  {\bibfnamefont {F.~A.}\ \bibnamefont {Zwanenburg}}, \bibinfo {author}
  {\bibfnamefont {D.~N.}\ \bibnamefont {Jamieson}}, \bibinfo {author}
  {\bibfnamefont {A.~S.}\ \bibnamefont {Dzurak}}, \ and\ \bibinfo {author}
  {\bibfnamefont {A.}~\bibnamefont {Morello}},\ }\bibfield  {title} {\enquote
  {\bibinfo {title} {{High-fidelity readout and control of a nuclear spin qubit
  in silicon}},}\ }\href {\doibase 10.1038/nature12011} {\bibfield  {journal}
  {\bibinfo  {journal} {Nature}\ }\textbf {\bibinfo {volume} {496}},\ \bibinfo
  {pages} {334--338} (\bibinfo {year} {2013})},\ \Eprint
  {http://arxiv.org/abs/1302.0047} {arXiv:1302.0047} \BibitemShut {NoStop}%
\end{thebibliography}%

\end{document}